\documentclass[useAMS,usenatbib]{mn2e}

\usepackage{graphicx}

\newcommand       \be           {\begin{equation}}
\newcommand       \ee           {\end{equation}}
\newcommand       \bea          {\begin{eqnarray}}
\newcommand       \eea          {\end{eqnarray}}
\newcommand       \apj          {ApJ}
\newcommand       \apjl         {ApJL}
\newcommand       \aap          {A\&A}
\newcommand       \nat          {Nature}
\newcommand       \mnras        {MNRAS}
\def\simlt{\mathrel{\hbox{\rlap{\hbox{\lower4pt\hbox{$\sim$}}}\hbox{$<$}}}}
\def\simgt{\mathrel{\hbox{\rlap{\hbox{\lower4pt\hbox{$\sim$}}}\hbox{$>$}}}}

\title[Magnetar Driven Bubbles and GRBs]{Magnetar Driven Bubbles and
the Origin of Collimated Outflows in Gamma-ray Bursts}
\author[N. Bucciantini, E. Quataert, J. Arons, B.~D. Metzger,
T.~A. Thompson]{N. Bucciantini$^{1}$\thanks{E-mail:
nbucciantini@astro.berkeley.edu}, E. Quataert$^{1}$, J. Arons$^{1,2,3}$,
B.~D. Metzger$^{1,2}$, T.~A. Thompson$^{4}$\\ 
$^{1}$Astronomy Department and Theoretical Astrophysics Center,\\
~~~~~~University of California at Berkeley, 601 Campbell Hall, Berkeley CA,
94720, USA\\
$^{2}$Physics Department, University of California at Berkeley \\
$^{3}$Kavli Institute for Particle Astrophysics and Cosmology, Stanford University\\
 $^{4}$Lyman Spitzer Jr.~Fellow, Department of Astrophysical Sciences, Peyton Hall,
Ivy Lane, Princeton, NJ, 08544, USA}
\begin{document}

\date{Accepted . Received ; in original form }

\pagerange{\pageref{firstpage}--\pageref{lastpage}} \pubyear{????}

\maketitle

\label{firstpage}

\begin{abstract}

We model the interaction between the wind from a newly formed rapidly
rotating magnetar and the surrounding supernova shock and host star.
The dynamics is modeled using the two-dimensional, axisymmetric
thin-shell equations. In the first $\sim 10-100$ seconds after core
collapse the magnetar inflates a bubble of plasma and magnetic fields
behind the supernova shock.  The bubble expands asymmetrically because
of the pinching effect of the toroidal magnetic field, even if the
host star is spherically symmetric, just as in the analogous problem
of the evolution of pulsar wind nebulae. The degree of asymmetry
depends on $E_{mag}/E_{tot}$, the ratio of the magnetic energy to the
total energy in the bubble.  The correct value of $E_{mag}/E_{tot}$ is
uncertain because of uncertainties in the conversion of magnetic
energy into kinetic energy at large radii in relativistic winds; we
argue, however, that bubbles inflated by newly formed magnetars are
likely to be significantly more magnetized than their pulsar
counterparts. We show that for a ratio of magnetic to total power
supplied by the central magnetar $\dot E_{mag}/\dot E_{tot} \simlt
0.1$ the bubble expands relatively spherically.  For $\dot
E_{mag}/\dot E_{tot} \simgt 0.3$, however, most of the pressure in the
bubble is exerted close to the rotation axis, driving a collimated
outflow out through the host star. This can account for the
collimation inferred from observations of long-duration gamma-ray
bursts (GRBs).  Outflows from magnetars become increasingly
magnetically dominated at late times, due to the decrease in
neutrino-driven mass loss as the young neutron star cools.  We thus
suggest that the magnetar-driven bubble initially expands relatively
spherically, enhancing the energy of the associated supernova, while at
late times it becomes progressively more collimated, producing the
GRB. The same physical processes may operate in more modestly rotating
neutron stars to produce asymmetric supernovae and lower energy
transients such as X-ray flashes.

\end{abstract}


\section{Introduction}

In the first few seconds after core collapse in a massive star, a
proto neutron star (PNS) cools and contracts on its Kelvin-Helmholz 
cooling timescale ($\sim 10-100$ s), radiating its
gravitational binding energy ($\sim10^{53}$ ergs) in neutrinos
(\citealt{burrows86}, \citealt{pons99}).  The cooling epoch is
accompanied by mass loss driven by neutrino energy deposition in the
atmosphere of the PNS (e.g. \citealt{ducan86}, \citealt{qian96},
\citealt{thom01}).

A subset of neutron stars are inferred to have magnetic field
strengths of $\sim 10^{14}-10^{15}$ G (``magnetars''; see \citealt
{woods04} for a review).  If some magnetars are born with millisecond
rotation periods (e.g., \citealt{duncan92}, \citealt{thom93}), the
combination of rapid rotation and strong magnetic fields makes the
winds from young magnetars significantly more energetic than the
thermally driven winds from slowly rotating neutron stars.  In
addition, as the neutrino-driven mass loss decreases during the
Kelvin-Helmholtz cooling epoch, the wind from a proto-magnetar becomes
increasingly magnetically-dominated and the flow eventually becomes
relativistic.  For this reason, proto-magnetars have been considered
as a possible central engine for long-duration gamma ray bursts
(GRBs)\footnote{In what follows we typically drop the phrase
``long-duration'' for conciseness and refer to long-duration GRBs
simply as GRBs.} and hyper-energetic supernovae (\citealt{usov92},
\citealt{thom94}, \citealt{wheeler00}, \citealt{thom04}), and as a
possible source of ultra-high energy cosmic rays (\citealt{blasi00},
\citealt{arons03}).

The discovery that GRBs are at cosmological distances confirmed that
the isotropic energy scale for the gamma-ray emission from GRBs is
$\sim 10^{52}-10^{53}$ ergs (see, e.g., \citealt{woo06}).  However,
the interpretation of afterglow observations (``jet breaks'')
suggested that GRBs are powered by collimated jets and that the
intrinsic energy in relativistic material is $\sim 10^{51}$ ergs
(e.g., \citealt{frail01}).  This interpretation has become less clear
in recent years because of the complex time-dependence in SWIFT X-ray
afterglow observations and the lack of evidence for X-ray jet breaks
in the first $\sim 10$ days (e.g., \citealt{sato07}, \citealt{rac07}).
Nonetheless, the case for collimated outflows from GRBs is compelling.
Theoretically, the association of many long-duration GRBs with
supernovae (\citealt{woo06}) sets the natural energy scale for GRBs at
$\sim 10^{51}-10^{52}$ ergs.  In addition, estimates of the energy in
relativistic outflows in GRBs from late time radio observations
provide lower limits of the same order, although the true energy could
in principle be much higher (see, e.g., \citealt{eic05}).

In the collapsar model (e.g., \citealt{mcf99}), the collimated
outflows from GRBs are accounted for by jets produced by an accretion
flow onto a central black hole.  In the magnetar model, the origin of
such collimated outflows is less clear.  Relativistic magnetized
outflows by themselves do not efficiently self-collimate (e.g.,
\citealt{le01}).  Although observations of pulsar wind nebulae
(PWNe)-- which are intrinsically far more relativistic than GRBs --
show jet-like features (e.g., \citealt{weiss00}, \citealt{pavlov01},
\citealt{gaensler02}), these are believed to be only mildly
relativistic outflows produced by the interaction between the pulsar
wind and the surrounding expanding supernova (SN) remnant
(\citealt{kom04}, \citealt{ldz04}).  In this paper, we explore the
hypothesis that collimated outflows from newly formed magnetars can
likewise be produced by the interaction between the magnetar wind and
the surrounding host star.  

Our physical picture is that the fast trans-magnetosonic magnetar wind
shocks on the relatively slow outgoing SN envelope, creating a
subsonic bubble of plasma and magnetic fields inside its host star.
Because of the strong toroidal magnetic field and the accompanying
pinch, an anisotropic pressure distribution between the pole and
equator is set up within the cavity defined by the SN shock and the
incoming magnetar wind.  For simplicity we assume that (1) an outgoing
SN shock has created a central evacuated cavity and (2) the
surrounding host star is spherically symmetric.  Assumption (1) allows
us to model the problem of interest as a free magnetar wind
interacting with the expanding envelope created by a SN shock that is
in turn sweeping through the host star.  Spectral modeling of the
hyper-energetic supernovae associated with several GRBs suggests
massive progenitor stars (e.g., \citealt{iwa98}; \citealt{maz06a}).
This has been interpreted as indicating that GRBs are associated with
the formation of black holes.  However, there is increasing evidence
that some Galactic magnetars arise from massive stars with ZAMS masses
of $\approx 40 M_\odot$ (e.g., \citealt{mun06}).  Thus our assumption
of a successful core-collapse SN leaving behind a rapidly rotating
magnetar is quite reasonable given current observational constraints
on the progenitors of magnetars and GRBs.  Our assumption (2) that the
host star is spherically symmetric may be conservative.
Multi-dimensional simulations of core-collapse in the presence of
rapid rotation and strong poloidal magnetic fields find that the
explosion may occur preferentially along the rotation axis (e.g.,
\citealt{leblanc70}, \citealt{burrows07}).  It is presumably easier to
produce a late-time collimated outflow in this case, since a low
pressure, low inertia channel has already been created.

A full magnetohydrodynamic (MHD) simulation of the interaction between
a magnetar wind and its host star would require resolving a very wide
radial dynamic range.  In addition, the physical conditions in the
wind at large distances -- in particular, the magnetization of the
wind -- are not fully understood (\S 2).  For these reasons, we
believe that it is fruitful to solve a model problem that allows one
to readily explore the parameter space of magnetar-host star
interactions -- the thin-shell approximation provides such a model.

In the thin-shell approximation, one assumes that the material
swept-up by the wind from the central object is confined to a
geometrically thin shell, whose dynamics is then evolved (e.g.,
\citealt{giu82}).  This model has been extensively applied in the
study of the interaction of stellar winds with their surrounding
environment, both in the case of momentum driven winds (see, e.g.,
\citealt{canto80,canto96,wilkin00}) and in the case of pressure driven
winds (e.g., \citealt{chev94}).  The evolution of magnetized PWNe
bounded by an expanding SN remnant (\citealt{beg92}) is the closest
analogue to the problem we consider in this paper.  In a number of
cases, more detailed numerical simulations have confirmed the validity
of the thin-shell model (see, e.g., \citealt{stev92,me02} for
hydrodynamical examples).  Most importantly for our purposes,
axisymmetric relativistic MHD simulations by \citet{van03} and
\citet{ldz04} have shown that the overall shape of PWNe 
resembles that predicted by the thin-shell model of
\citet{beg92}.  For these reasons we believe that the thin-shell shell
approximation is a useful tool for studying the structure and
evolution of bubbles formed by magnetar winds inside their progenitor
stars.  In addition, these calculations can define the most interesting
parameter space for future relativistic MHD simulations.

The remainder of this paper is organized as follows.  In \S~2 we
discuss the general properties of proto-magnetar winds, and how they
evolve in the $\sim 100$ seconds after core-collapse. We also discuss
the equilibrium structure of the magnetized bubble created by the
magnetar wind behind the SN shock. Section~3 summarizes the thin-shell
equations. In \S~4 we present our results for the evolution of the SN
shock due to the asymmetric pressure produced by the interior
magnetized bubble.  In \S~5 we summarize our conclusions and discuss
the implications of our results for understanding observations of
long-duration gamma-ray bursts, X-ray flashes, and asymmetric
supernovae.  In the Appendix we present self-similar solutions that
provide insight into how the shape of the bubble is related to its
magnetization and the conditions in the ambient medium.

\section[]{Protomagnetar Evolution and Bubble Structure}

\begin{figure}
\begin{center}
\includegraphics[scale=0.5]{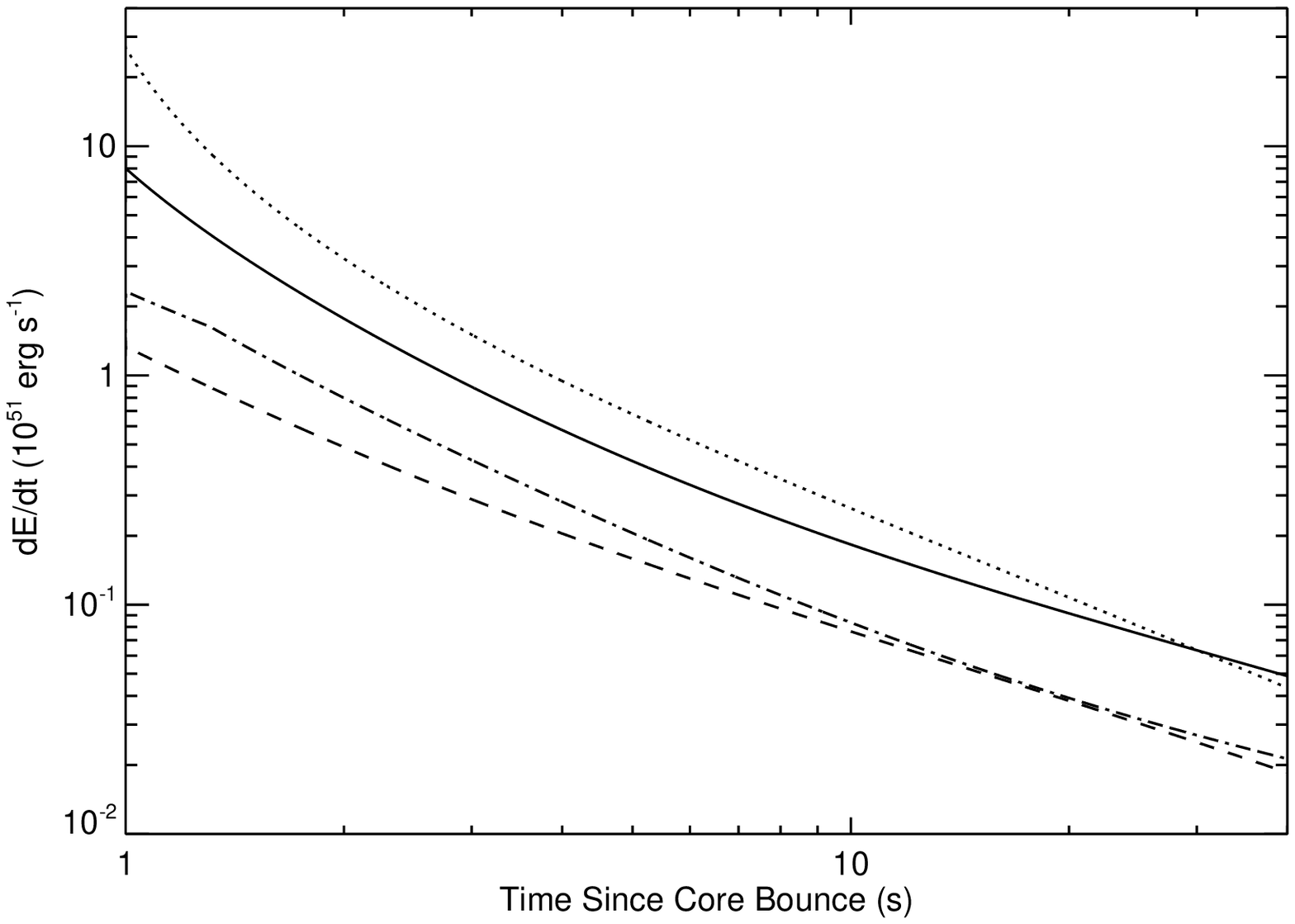}
\includegraphics[scale=0.5]{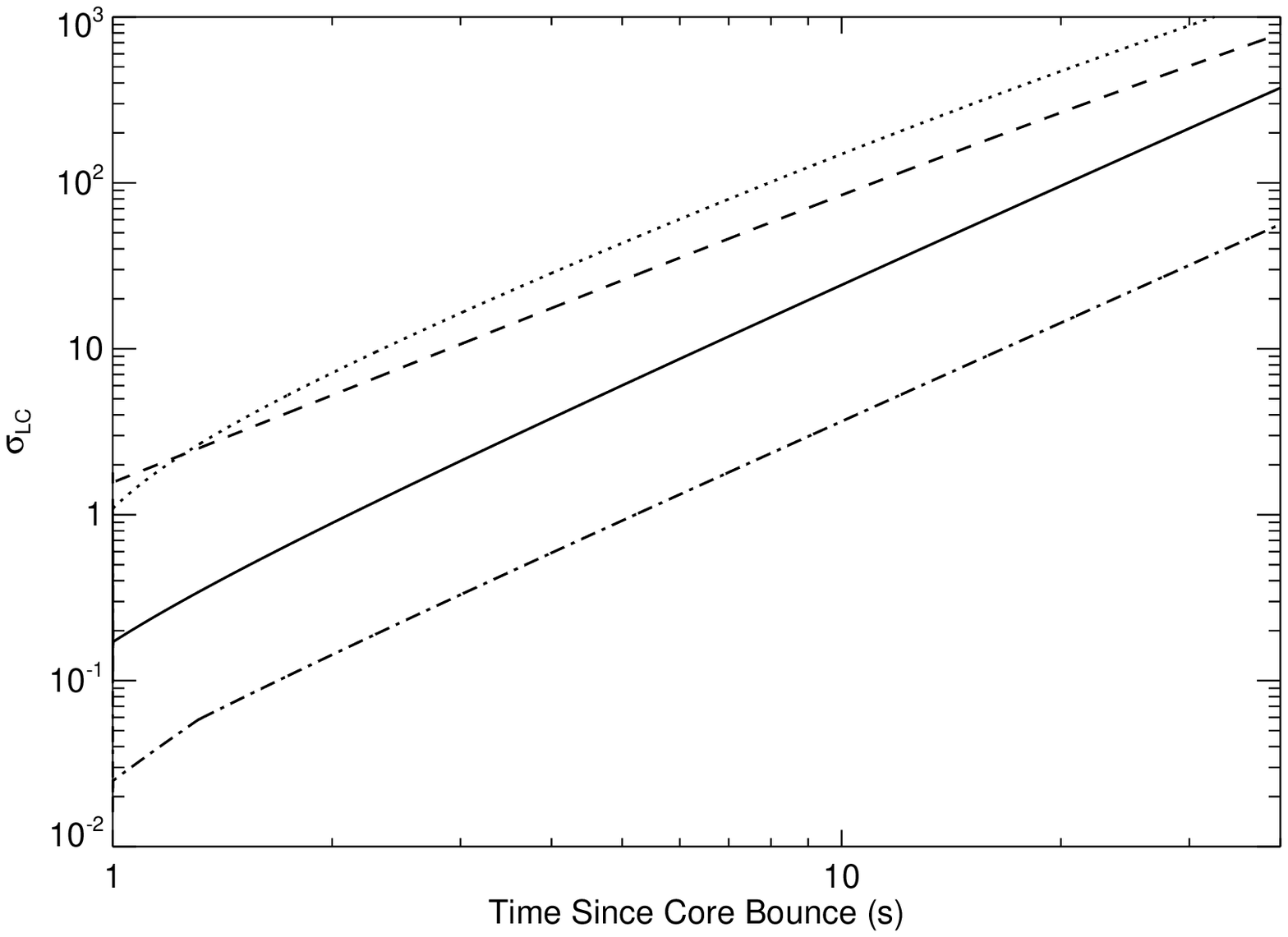}
\caption{Evolution of a magnetar wind as a function of time since core
bounce, based on the evolutionary calculations of \citet{met07}.  {\it Top}: Energy loss rate {\it Bottom}: Magnetization
at the light cylinder.  Models are for an initial period of P = 1 ms,
$B = 10^{15}$ G (dot dashed), $B = 3 \times 10^{15}$ G (solid), and $B
= 10^{16}$ G (dotted) and an initial period of $P = 2$ ms and $B =
10^{16}$ G (dashed).}
\label{fig:wind}
\end{center}
\end{figure}


\subsection{Protomagnetar Wind Evolution \label{sec:windevol} }

In order to set the stage for the thin-shell evolutionary calculations
that follow, this section summarizes some of the key properties of
outflows from young rapidly rotating magnetars. \citet{met07}
(hereafter M07) solved the one-dimensional (monopolar) neutrino-heated
non-relativistic MHD wind problem for conditions appropriate to young
magnetars.  These calculations provide the mass-loss rate ($\dot M$)
and energy-loss rate ($\dot E$) from the magnetar as a function of
parameters such as the neutrino luminosity, magnetic field strength
$B$, and rotation rate $\Omega$.  The calculation of $\dot M$ is
applicable even if the wind is relativistic because the mass-loss is
set in the non-relativistic subsonic portion of the wind at small
radii.  The calculations of M07 include the neutrino micro-physics
necessary for direct application to PNSs.  Their calculations were,
however, restricted to monopolar field structure.  A complementary set
of calculations was carried out by \citet{me06} (hereafter B06), who
studied aligned dipolar (axisymmetric) non-relativistic and
relativistic MHD winds from neutron stars assuming an adiabatic
equation of state.  M07 showed that their results could be combined
with those of B06 to provide evolutionary models for early magnetar
spin-down, including the transition from non-relativistic to
relativistic outflows as the neutrino-driven mass loss decreases.

Figure \ref{fig:wind} shows the energy loss rate $\dot E$ and
magnetization $\sigma_{LC}$ as a function of time from the
evolutionary calculations of M07 for several combinations of (dipole)
magnetic field strengths and magnetar birth period.  The values of $B
\approx 10^{15}-10^{16}$ G and $P \approx 1-2$ ms are chosen to be
characteristic of PNSs that produce conditions suitable for producing
GRBs or hyper-energetic supernovae.
 The magnetization in Figure \ref{fig:wind} is
defined by \be \sigma_{LC} \equiv \Phi_{B}^{2}\Omega^{2}/\dot{M}c^{3}
\label{sig}, \ee where $\Phi_{B}$ is the total open magnetic flux per
$4\pi$ steradian (\citealt{mic69}), $\dot M$ is the mass loss rate,
and $\sigma_{LC}$ is evaluated at the light cylinder.  Winds with
$\sigma_{LC} \simlt 1$ are non-relativistic while those with
$\sigma_{LC} \simgt 1$ are relativistic.  The calculations in Figure
\ref{fig:wind} assume that the PNS luminosity decreases in time
$\propto t^{-1}$ until $t = 40$ s, motivated by the cooling evolution
of \citet{pons99}.  We note, however, that the cooling of strongly
magnetized rapidly rotating PNSs is not well understood, which is a
source of uncertainty in the determination of $\dot M(t)$, $\dot E(t)$
and $\sigma_{LC}(t)$.  

The energy loss rates $\dot E$ in Figure \ref{fig:wind} are a factor
of $\sim 10$ larger than the ``vacuum dipole'' rate for the given
value of $\Omega$ and $B$.  There are two reasons for this.  First,
when the wind is non-relativistic ($\sigma_{LC} \simlt 1$), the energy
loss rate is larger than in the relativistic limit, with $\dot E
\propto \dot M^{1/3} \, (\dot M^{3/5})$ for non-relativistic
magnetized winds with a monopole (dipole) magnetic field structure
(e.g., \citealt{thom04}). In addition, the large mass loss rates
accompanying PNS spin-down lead to excess open magnetic flux which
enhances the spin-down of the PNS (see the simulations of B06).  This
is true even when the wind is moderately relativistic ($\sigma_{LC}
\simgt 1$).  The large energy loss rates shown in Figure
\ref{fig:wind} are sufficient to lead to appreciable spin-down of the
PNS during the Kelvin-Helmholtz epoch.  For example, for the model with
$P = 1$ ms and $B = 3 \times 10^{15}$ G in Figure \ref{fig:wind}
(solid line), the PNS loses $\approx 80\%$ of its rotational energy in
the first 40 seconds.  This efficient spin-down is largely responsible
for the fact that $\dot E$ decreases in time as the PNS cools (see
Figure \ref{fig:wind}).\footnote{Two additional effects contribute to
the decline in $\dot E$ with time.  First, as the PNS cools, the mass
loss rate $\dot M$ decreases.  In the non-relativistic limit, the
energy loss rate is proportional to $\dot M^{0.3-0.6}$ and thus
decreases as well (this is relevant for the $P = 1$ ms, $B = 3 \times
10^{15}$ (solid line) and $P = 1$ ms, $B = 10^{15}$ G (dot-dashed)
models in Figure \ref{fig:wind} at early times).  The decreasing mass
loss rate also decreases the fraction of open magnetic flux and thus
$\dot E$.}

As the PNS cools, the neutrino-driven mass loss decreases in time.
This in turn causes a transition from a non-relativistic to
relativistic wind, as shown explicitly in the plot of $\sigma_{LC}(t)$
in Figure \ref{fig:wind}.  These calculations of $\sigma_{LC}$ are
based on equatorial spin-down models (M07), which probably
underestimate the angle-averaged $\sigma$ in the wind by a factor of
few (B06).  Nonetheless, the evolution from a moderately mass-loaded
marginally relativistic wind ($\sigma_{LC} \sim 1$) to a highly
magnetized Poynting flux dominated outflow ($\sigma_{LC} \gg 1$) is
expected to be generic for cooling magnetars.

As we show in the next section, the impact of the magnetar on its host
star depends critically on the strength of the magnetic field in the
bubble created by the magnetar wind; the generation rate of the field 
in the bubble is in turn determined by the
magnetization $\sigma$ of the wind at large radii.  In
non-relativistic winds, the magnetic energy and kinetic energy are in
approximate equipartition at large radii, with $E_{mag} \approx 2
E_{kin}$ (e.g., \citealt{lam99}).  One-dimensional models of ideal
relativistic winds, however, find that the asymptotic Lorentz factor
of the wind is $\gamma_\infty \approx \sigma_{LC}^{1/3}$ and the
asymptotic magnetization is $\sigma \approx \sigma_{LC}^{2/3}$
(\citealt{mic69}, \citealt{gol70}) so that most of the energy remains
in the magnetic field at large radii.  These results apply in the
limit of $\sigma_{LC} \gg 1$. Relativistic MHD simulations
(\citealt{me07}) show that for intermediate values of
$\sigma_{LC}\simlt 20$, a reasonable fraction of the magnetic energy
is converted into kinetic energy at large distances, with rough
equipartition obtaining by $\sim 10^4$ stellar radii.

In the limit of very high $\sigma_{LC}$, studies of PWNe (e.g. the
Crab Pulsar) find that the wind must have low $\sigma \sim 10^{-2}$ at
large radii (e.g., \citealt{ken84}, \citealt{beg92}).  Although there
is no consensus on the mechanism responsible for the inferred decrease
in pulsar wind magnetization at large radii, a prominent class of
models relies on magnetic dissipation in the relativistic outflow over
a large radial distance (e.g., \citealt{cor90}; \citealt{le01b};
\citealt{kir03}).  The physical conditions in proto-magnetar winds are
quite different from those in pulsar winds (e.g., they are much denser
so that there is no charge starvation).  In addition, the distance to
the termination shock is much smaller in the SN confined winds from 
young magnetars, $\sim 10$
light cylinder radii (see below) compared to more than $10^4$ light
cylinder radii in PWNe and in pulsar-Be star binaries.  The reduced
flow time between the light cylinder and the termination shock may
mean that dissipation of magnetic energy in young magnetar winds is
less complete than in pulsar winds. As a result, we suspect that the
rate of injection of magnetic energy into bubbles created by
protomagnetars may be significantly larger than that inferred in the
PWNe context.  Given the uncertainties, however, we treat the
magnetization in the outflow, expressed as the ratio of the magnetic
energy injection to the total power ($\dot E_{mag}/\dot E_{tot}$), as
a free parameter in this paper, bearing in mind the generic evolution
from $\sigma_{LC} \sim 1$ to $\sigma_{LC} \gg 1$ in Figure
\ref{fig:wind}.

The models shown in Figure \ref{fig:wind} assume that the wind from
the central magnetar is freely expanding into a cavity evacuated by
the outgoing SN shock.  Formally, this requires that the radius of the
fast magnetosonic point must be smaller than the radius of the SN
shock; the latter is $R_s \sim 10^9$ cm in the first few seconds,
which is indeed larger than the typical distance to the fast surface of $\sim
10-40$ neutron star radii (B06, for a millisecond rotator). 
 As the freely expanding wind moves
out, it interacts with the surrounding SN shock and previously shocked
wind material.  More precisely, the wind will reach a termination
shock at which its kinetic energy is thermalized and the magnetic
field is compressed.  A correct determination of the size of the
termination shock requires a full MHD model of the wind-bubble
interaction (e.g., \citealt{ldz04}).  As a rough guide to the relevant
scales, however, we note that in the simple case of a constant $\dot
M$ and $\dot E$, $\sigma_{LC} \sim 1$ wind moving into a spherically
symmetric bubble, the termination shock is located at a radius $R_t
\sim R_s (R_s/ct)^{1/2} \sim 0.1 R_s \sim 10^8$ cm where $t$ is the
time since the onset of the wind (in sec).  For $R_t < R < R_s$, the
wind develops into a bubble of plasma and magnetic field confined by
the SN shock and host star.

\subsection{The Bubble Structure}

\label{sec:bubble}

If one neglects plasma flow inside the bubble, a simple solution for
the structure inside the bubble ($R_t < R < R_s$) can be obtained in
the case of the predominantly toroidal magnetic field expected at
large radii in the wind.  This solution was found by \cite{beg92}.  We
reproduce several of its features here because they are important to
our model. 

The \cite{beg92} solution will be valid as long as typical flow speeds
do not exceed the local sound speed.  In the case of a relativistic
magnetized bubble the sound speed ranges from $c/\sqrt{3}$ to $c$.  It
is possible that, close to the termination shock, post shock flow can
move with high velocities \citep{ldz04}, but in the bulk of the
bubble, typical speeds are expected to be a small fraction of $c$,
unless the cavity itself expands at a considerable fraction of the
speed of light.  Indeed, as long as the expansion velocity of the
shell is small compared to the sound speed inside the bubble, the
plasma inside will always relax to pressure equilibrium, independent
of the energy distribution in the wind (be it primarily polar as for a
non-relativistic wind or primarily equatorial as for a relativistic
wind).  Neglecting the fluid flow, the structure is given by the
assumption of magnetohydrostatic equilibrium.  Assuming axisymmetry,
the momentum equations become:
\begin{eqnarray}
\frac{\partial}{\partial z}\left(p+\frac{B^2}{8\pi}\right)=0,\;\;\;
\frac{\partial}{\partial r}\left(p+\frac{B^2}{8\pi}\right)+\frac{B^2}{4\pi r}=0,
\label{hydro}
\end{eqnarray}
where $r$ is the cylindrical radius, $p$ is the pressure, and $B$ the
toroidal magnetic field in the bubble.  The first equation simply
states that isobaric surfaces are coaxial cylinders. 

If entropy is constant along each flow streamline in the bubble then
the continuity equation can be written as:
\begin{equation}
\frac{\partial}{\partial r}(p^{1/\Gamma}rv_r)+\frac{\partial}{\partial z}(p^{1/\Gamma}rv_z)=0.
\end{equation}
where $\Gamma$ is the adiabatic index of the fluid.  Comparing
this with the flux-freezing condition for the toroidal magnetic field yields
\begin{equation}
p\propto(B/r)^{\Gamma}.
\label{flux}
\end{equation}
For the case of a relativistic plasma ($\Gamma = 4/3$), equation
(\ref{flux}) can be used in the r-momentum equation to find
\begin{equation}
p=\frac{p_n}{\zeta^2},\;\;\;\frac{B^2}{8\pi}=\frac{9p_nr^2}{16\zeta^3 H^2},
\label{p}
\end{equation}
where $\zeta$ is the solution of the following equation:
\begin{equation}
(\zeta+\frac{9r^2}{32H^2})^2-\zeta^3=0.
\label{zeta}
\end{equation}

\begin{figure}
\resizebox{\hsize}{!}{\includegraphics[bb=100 366 550 715,clip]{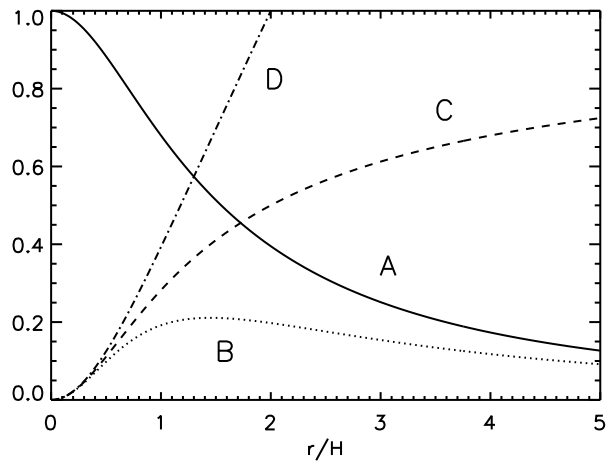}}
\caption{Pressure structure for a relativistic plasma containing
toroidal magnetic field in magneto-hydrostatic equilibrium (based on
\citealt{beg92}). The solid line (A) is the total pressure, normalized to
the value on the axis. The dotted line (B) is the magnetic pressure,
normalized to the total pressure on the axis. The dashed line (C) is the
ratio of the magnetic to the total pressure, while the dash-dotted
line (D) is the ratio of the magnetic to the thermal pressure. Bubbles
with weak magnetic fields have large values of $H$ relative to the
size of the bubble (see Figure \ref{fig:H}) and thus only the $r \ll H$
part of this plot is applicable: as a result the pressure is relative
uniform and the system will expand spherically.  By contrast, bubbles
with appreciable magnetic fields have smaller values of $H$ and thus
the pressure on the axis is significantly larger than the pressure
near the equator.  Such bubbles will expand asymmetrically.}
\label{fig:beg}
\end{figure}

The solution for the pressure in the bubble given by equations
(\ref{hydro})-(\ref{zeta}) depends on two parameters.  One of these,
the pressure on the axis $p_n$, determines the overall magnitude of
the pressure in the bubble.  The other, the scale height $H$ of the
pressure distribution, determines the pressure stratification in the
bubble.  In Figure~\ref{fig:beg} we plot the normalized pressure
profile derived from the solution of the above equations. The total
pressure is higher along the axis ($r=0$) and asymptotically decreases
as $r^{-2}$. The region close to the axis contains a low $\sigma$
plasma and is essentially pressure dominated, while at larger
distances the plasma is magnetically dominated, and the ratio of
magnetic to thermal pressure increases linearly with the distance.
Equipartition is reached for $r/H\sim 2$. The results in Figure
\ref{fig:beg} assume a relativistic plasma with $\Gamma = 4/3$, which
corresponds to $\sigma_{LC} \simgt 1$ in Figure \ref{fig:wind}. The
magnetar wind may be non-relativistic at very early times, so that
$\Gamma = 5/3$ is more appropriate. For $\Gamma = 5/3$ the pressure
profiles are qualitatively similar to those in Figure \ref{fig:beg},
although the scale height $H$ is a factor of $\approx 2$ smaller for a
given ratio of magnetic to total energy in the bubble.  For
simplicity, we simply set $\Gamma = 4/3$ in all of our calculations.

\begin{figure}
\resizebox{\hsize}{!}{\includegraphics[bb=111 360 555 720,clip]{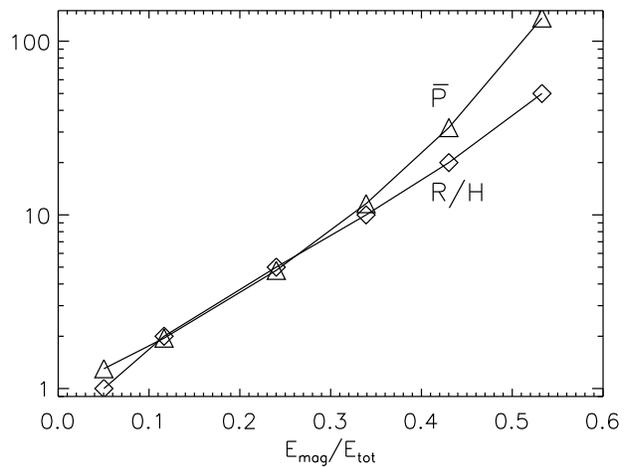}}
\caption{Dimensionless pressure $\bar P$ (see eq. [\ref{paxis}]) and
scale-height $H$ as a function of the magnetic energy in the bubble,
for the case of spherical bubble. $\bar P$ is the ratio of the
pressure on axis to the pressure in an unmagnetized spherical
bubble. For even modest magnetic energy, the pressure distribution
becomes highly anisotropic with $\bar P \gg 1$ and $H \ll R$.}
\label{fig:H}
\end{figure}

The scale height $H$ and the asymmetry of the pressure distribution can
be expressed in terms of the ratio of the magnetic energy to total
energy in the bubble.  To quantify this effect, consider a spherical
bubble of radius $R$ and total energy $E$.  The pressure along the
axis is given by 
\be 
p_n \approx 8 \times 10^{22} \, \bar P \left(E
\over 10^{51} \, {\rm ergs} \right) \left (R \over 10^9 \, {\rm cm}
\right)^{-3} \, {\rm ergs \, cm^{-3}}. \label{paxis} 
\ee 
The dimensionless number $\bar P$ is the pressure on the axis relative to
that in an unmagnetized bubble.  Figure \ref{fig:H} shows $\bar P$ and
the scale height $H/R$ as a function of $E_{mag}/E_{tot}$, the ratio
of the magnetic to total energy in the bubble (similar results are
obtained for the self-similar solutions described in the Appendix;
see, e.g., Figure \ref{fig:andam1}).  Magnetized bubbles have $\bar P
\gg 1$ and $H \ll R$ (where $R$ is the radius of the bubble, not the
cylindrical radius within the bubble used above and in Figure
\ref{fig:beg}).  Figure \ref{fig:beg} shows that, due to the pinching effect of the toroidal magnetic field, the pressure in the
bubble in this case will be concentrated along the axis and so the
bubble will expand asymmetrically.  By contrast, very weakly
magnetized bubbles have $H \simgt R$ and roughly constant pressure
throughout.  Note that a magnetization of $E_{mag}/E_{tot} \simgt 0.1$
is required to make $H \simlt R$ and the pressure distribution in the
bubble relatively asymmetric. 

We now calculate how the swept-up shell in the host star responds to the
pressure produced by the magnetized bubble created by the central
magnetar.

\section[]{The Thin-Shell Equations}

The equations describing the evolution of an axisymmetric bubble can
be greatly simplified if one assumes that the swept-up material is
confined in a thin-shell, so that the dynamics can be described in the
``thin-shell approximation.'' This approximation is valid as long as
the thickness of the shell is small compared to the radius of
curvature.  The thin-shell equations
account for conservation of mass and momentum.  A detailed derivation
of the equations can be found in \citet{giu82} where corrections for the
thickness of the shell are also discussed. In the case of
infinitesimally thin shell they are given by:
\begin{eqnarray}
\tan{\xi}=-\frac{1}{R}\frac{\partial R}{\partial \theta},\label{eq:1}\\
u_{\bot}=\cos{\xi}\frac{\partial R}{\partial t},\\ 
u_{\|}=\sin{\xi}\frac{\partial R}{\partial t},\\
\frac{\partial (A\sigma)}{\partial t}=-\rho_o(v_{\perp o}-u_{\perp})A+\rho_i(v_{\bot i}-u_{\bot})A-\nonumber\\
\frac{\partial }{\partial \theta}\left[R\sin{\theta}\sigma(v_{\|}-u_{\|})\right],\label{eq:dens}\\
\frac{\partial (A\sigma {\bf v})}{\partial t}=-[\rho_o(v_{\bot o}-u_{\bot}){\bf v}_{o}+{\bf e}_{\bot}(p_o+B_o^2/8\pi)]A\nonumber\\+
 [\rho_i(v_{\bot i}-u_{\bot}){\bf v}_{i}+{\bf e}_{\bot}(p_i+B_i^2/8\pi)]A-\nonumber\\
\frac{\partial }{\partial \theta}\left[R\sin{\theta}\sigma(v_{\|}-u_{\|}){\bf v}\right],\label{eq:mom}\\
A=\left(\frac{R^2\sin{\theta}}{\cos{\xi}}\right).\label{eq:6}
\end{eqnarray}
where $\xi$ is the angle between the radial direction and the normal
to the shell surface, $A$ is the effective area of each element of the
shell, and $\sigma$ is the surface density. The suffixes $\bot$ and
$\|$ represent the directions perpendicular and parallel to the shell
(and they are time dependent), while $i$ and $o$ label the conditions
at the inner and outer edge of the shell. The velocity $v_\bot=u_\bot$
is the expansion velocity of the shell perpendicular to itself, $u_\|$
is the expansion velocity parallel to itself, and $v_\|$ is the flow
velocity along the shell.

Equation (\ref{eq:dens}) represents conservation of mass along the
shell, while equation (\ref{eq:mom}) describes momentum conservation
in the shell. Both equations include a flux of the relevant
quantity along the shell itself, and source terms due to the inner and
outer media. As discussed in \citet{giu82}, these equations employ a
Lagrangian remapping along the shell, and can be applied only as long
as $R(\theta)$ is a single valued function of $\theta$.

The evolution of the thin shell depends on the force driving it
(``i'') and on the conditions in the external medium (``o'').  The
solution in \S \ref{sec:bubble} describes the inner conditions used in
this study.  In many cases of interest, the outer medium is
sufficiently cold and the magnetic field is sufficiently weak that
their contribution to the pressure term in equation (\ref{eq:mom}) can
be neglected.  In our case, the outer medium is the outer part of the
progenitor star from which the magnetar formed (see \S \ref{bottle});
we do not neglect the thermal pressure, but we do assume that the
progenitor is unmagnetized.


Given the evolution of $H$ and $p_n$ with time (calculated below),
equations (\ref{eq:1})-(\ref{eq:6}) were solved under the assumption
of axisymmetry, to determine the evolution of the shape of the shell
with time.  We were not able to cast these equations in full upwind
form, because of the presence of a term describing the advection of
the shell curvature, which is not constant and changes in time. This
requires adding some artificial viscosity in order to avoid the
numerical growth of perturbations.

One cautionary comment about the thin-shell model is in
order. Equations (\ref{eq:1})-(\ref{eq:6}) are momentum conserving,
not energy conserving, in the sense that a shell expanding into an
ambient medium has constant momentum and thus its energy decreases in
time.  The equations do conserve energy, however, in the sense that
the work done by the interior bubble is self-consistently supplied to
the shell (see \S~\ref{bottle}), but some of this energy is then lost
as the shell expands and sweeps out into the ambient
medium. Unfortunately, it is not possible to conserve both momentum
and energy in the time dependent, thin-shell approximation (by
contrast, in the self similar case discussed in the Appendix, one can
satisfy both requirements, but in this case the time evolution is
factored out of the equations).  One consequence of this is that the
calculations that follow probably evolve somewhat more slowly than
would a true SN shock expanding into its host star, although we are
confident that our conclusions about generating asymmetric bubbles are
robust (e.g., the self-similar solutions in the Appendix show similar
asymmetry).

\section{A Magnetar in a Bottle}
\label{bottle}

In this section we use the thin-shell model to calculate the evolution
of the magnetized bubble inflated by a central magnetar.  As was
pointed out in \S2.1, one needs to derive the internal pressure
distribution in the bubble in order to solve for the dynamics of the
shell. In particular, one needs to know the value $p_n$ of the total
pressure on the axis and the value of the scale height $H$ of the
pressure distribution. Once these two parameters are known it is
possible to derive the pressure throughout the bubble, in particular
its value at the inner edge of the shell. One can show that given the
shape of the shell bounding the bubble, $p_n$, and $H$, the total
energy $E_{tot}$, the magnetic energy $E_{mag}$, and the magnetic flux
$\Phi$ inside the bubble itself are uniquely defined, where \bea
E_{tot}=\int_{V}p_n{\cal F}(r/H,z)dv,\label{eq:etot}\\
E_{mag}=\int_{V}p_n{\cal G}(r/H,z)dv,\label{eq:emag}\\
\Phi=\int_{A}\sqrt{p_n}{\cal Q}(r/H,z)da,\label{eq:fmag} \eea and
where $V$ is the volume of the bubble and $A$ is the area in the $r-z$
plane, delimited by the shell. The dimensionless functions ${\cal F,
G, Q}$ are given in terms of cylindrical coordinates, and can be
derived from the pressure and magnetic field given by equations
(\ref{p})-(\ref{zeta}) in \S~2.1.

In order to compute the evolution of the internal structure in the
bubble we subdivided each time step ($dt$) of the shell evolution into two
sub-steps. In the first sub-step, that we call adiabatic, we neglect
injection of energy and magnetic field by the central source, and we
compute the adiabatic losses due to expansion according to:
\begin{equation}
dE_{tot}=\int_Sp\;dV,
\end{equation}
where $p$ is the total pressure along the shell surface $S$ and $dV$
is the volume increment that results from the evolution of the shell
surface. Once the adiabatic losses are known one can derive the new
value for the total energy in the bubble. During this adiabatic step
the magnetic flux remains constant.  After the adiabatic step, the new
values of $p_n$ and $H$ are re-evaluated by solving the following equations:
\bea
E_{tot,a}=E_{tot}-dE_{tot}=\int_{V}p_n{\cal F}(r/H,z)dv,\\
\Phi=\int_{A}\sqrt{p_n}{\cal Q}(r/H,z)da,
\eea
where the integrals are computed using the values of $V$ and $A$ 
after the expansion. Using the new values of $p_n$ and $H$, 
we need to recompute the new magnetic energy inside the bubble $E_{mag,a}$,
because adiabatic losses act on the total energy. This is done 
using equation (\ref{eq:emag}).

In the second sub-step, that we call the injection step, the shape of
the bubble is assumed to be fixed and we compute the new values of the
total energy and the magnetic energy given the rate of total energy
and magnetic energy injection by the central magnetar. The two equations 
to be solved for $p_n$ and $H$ are:
\bea
E_{tot,a}+\dot E_{tot}dt=\int_{V}p_n{\cal F}(r/H,z)dv,\\
E_{mag,a}+\dot E_{mag}dt=\int_{V}p_n{\cal G}(r/H,z)dv,
\eea
and once $p_n$ and $H$ are known we can also recompute the magnetic 
flux $\Phi$, which will be needed in the next time step.
With this method we determine the evolution of the pressure
on the inner edge of the shell as a function of time given $\dot
E_{tot}(t)$ and $\dot E_{mag}(t)$ (by, e.g., the results of
Figure \ref{fig:wind}).

Based on modeling the spectra of supernovae associated with nearby
GRBs, there are some indications that GRBs arise from very massive
stars with ZAMS masses of $M \approx 40 M_\odot$ (e.g., \citealt{iwa98}).
 There are also observational indications that Galactic
magnetars are formed from comparably massive stars (\citealt{gae05},
\citealt{mun06}).  We thus consider the evolution of a magnetized
bubble inside a progenitor star of $35\;M_{\odot}$, using the
progenitor models of \citet{woo02}. We
have also considered lower progenitor masses down to $\approx
11\,M_{\odot}$, which may be more appropriate for the progenitors of X-ray 
flashes (\citealt{maz06a}).  We find little difference in the results for
different progenitors, at the level of quantitative detail to which
we aspire.  The most significant effect is that for fixed energy injection,
the bubble expands more slowly for more massive progenitors.
The relative insensitivity to progenitor mass can in part can be
understood by noting that the self-similar solutions described in the
Appendix show explicitly that the elongation of the bubble depends
only weakly on the density profile of the ambient medium.

As discussed in \S 2, our model of the magnetar wind assumes that it
is expanding into a cavity evacuated by the outgoing SN shock. To
initialize our simulations, we thus carve out a spherical cavity with
a radius of $10^9$ cm inside our progenitor, corresponding to the
region of infall in the first $\sim 1$ sec. We assume that this cavity
is bounded by a thin shell whose mass is equal to the mass that
originally was in the cavity region minus $1.4M_\odot$ (the canonical
mass for a neutron star). In all of our simulations, time is defined
after core bounce and the simulation starts 1 second after core
bounce.  Moreover we impart to the shell an outward velocity so that
the total shell energy at the beginning is $10^{51}$ ergs, enough to
trigger a SN. If instead one assumes an initially stationary shell,
the evolution is essentially unchanged for weakly magnetized bubbles
because the pressure of the bubble is relatively isotropic (this
assumes that the magnetar wind extracts at least $\sim 10^{51}$ ergs
at early times, as is the case in the models shown in Figure
\ref{fig:wind}).  For strong magnetization, the elongation of the
bubble along the axis is also nearly independent of the initial shell
energy.  However, for large $E_{mag}/E_{tot}$, the pressure in the
bubble near the equator can be so small that infall cannot be
prevented.  To model this case, a full hydrodynamic solution is
required.

We follow the evolution of the shell and interior bubble to large
distances, into the hydrogen envelope of the progenitor. For GRB
progenitors, the hot plasma confined inside will emerge into the
circumstellar medium once the shell surface reaches the outer edge of
the helium core. The initial material that emerges will probably only
move with modest Lorentz factor.  Subsequent material will, however,
rapidly accelerate through the channel carved by the magnetar wind,
reaching asymptotic Lorentz factors set roughly by the enthalpy of the
material in the bubble (assuming that $E_{mag} \simlt E_{thermal}$ in
the bubble).  This phase of evolution cannot be studied using the thin
shell-approximation, but requires full relativistic MHD simulations.
Nonetheless, it appears natural that a highly relativistic and
collimated outflow will emerge out of the cavity carved by the early
magnetized bubble.

\subsection{Results}

\begin{figure*}
\resizebox{\hsize}{!}{\includegraphics[bb=84 376 380
720,clip]{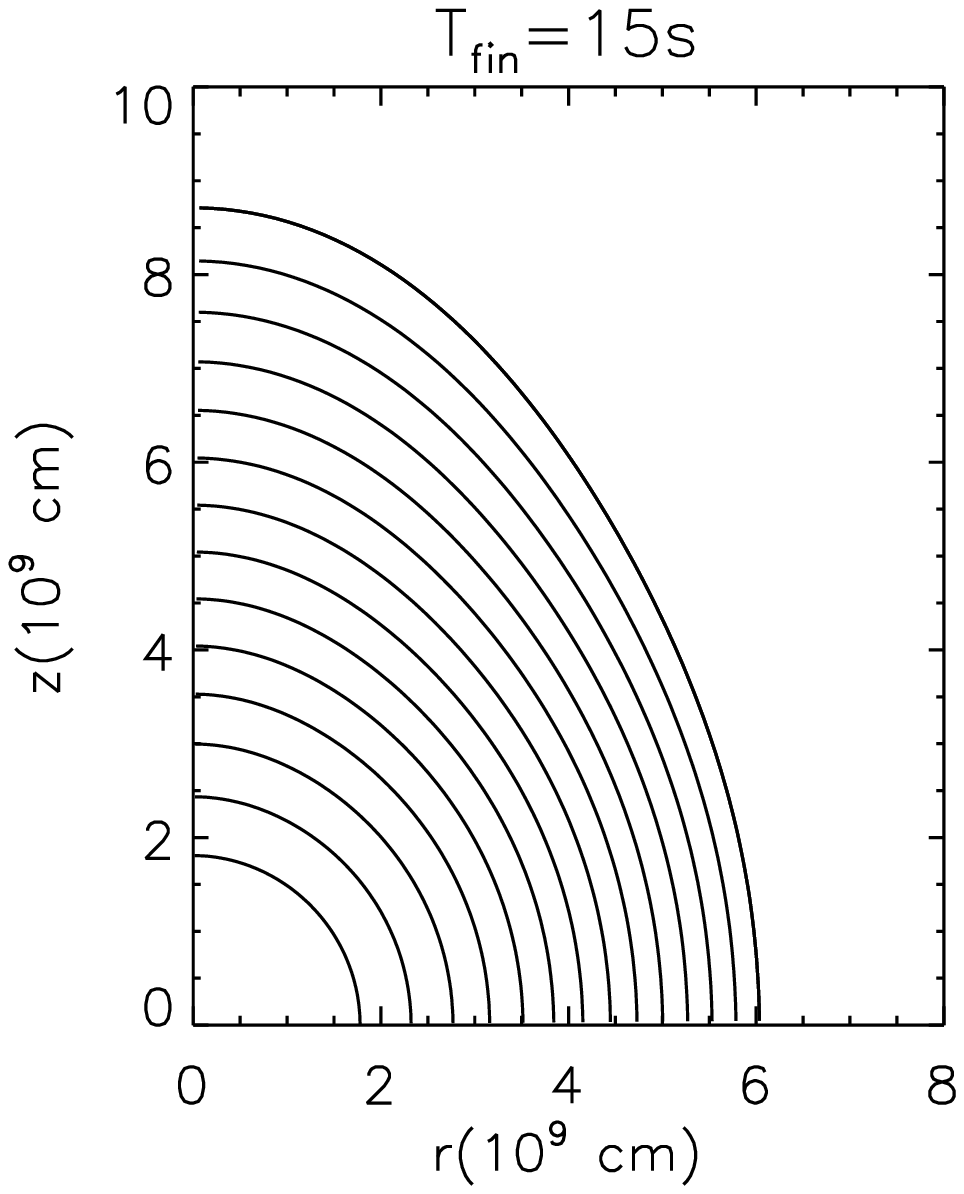}\includegraphics[bb=84 376 300
720,clip]{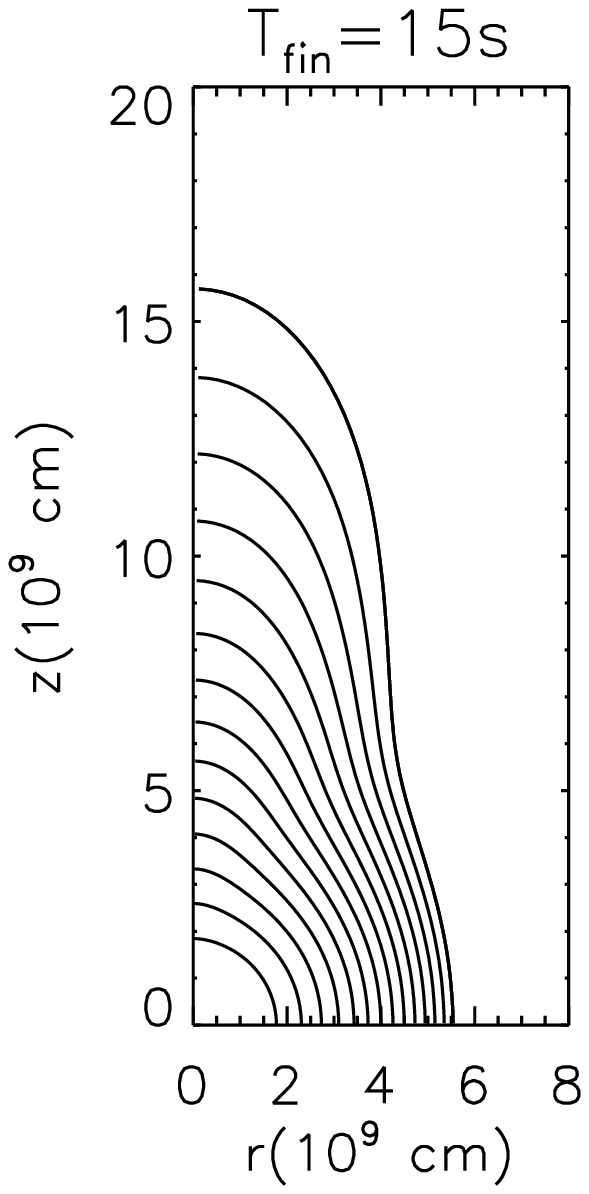}\includegraphics[bb=84 376 300
720,clip]{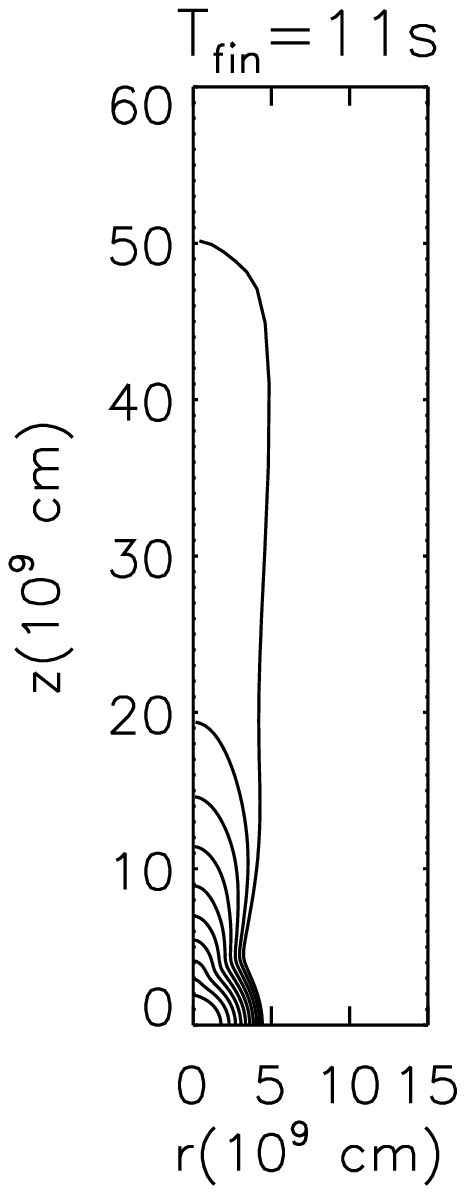}\includegraphics[bb=84 376 300 720,clip]{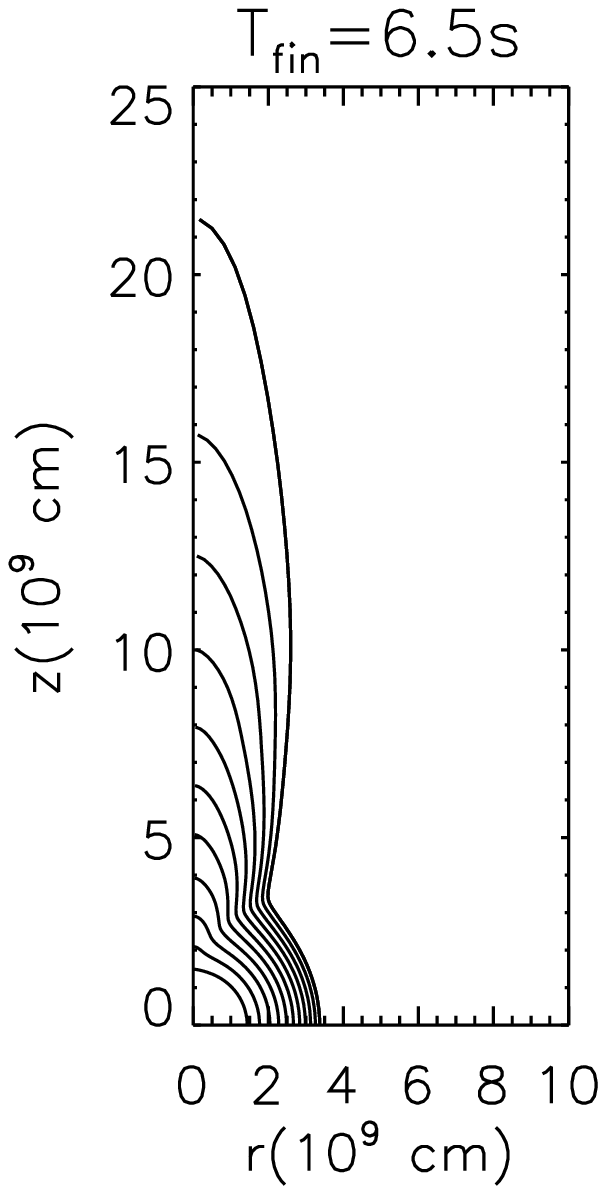}}
\caption{Evolution of a magnetized bubble inside a $35 M_\odot$
progenitor for a central source with $\dot E_{tot} = 10^{51} \, (t/1
\, {\rm s})^{-1} \, {\rm ergs \, s^{-1}}$. The initial time is 1s; the
contours describe the shape of the shell at 1s time intervals for the 
first three panels and 0.5s time intervals for the last. From
left to right, the solutions correspond to increasing the
magnetization of the bubble, with $\dot E_{mag}/ \dot
E_{tot}=0.1,0.2,0.3,\, \& \, 0.4$, respectively.  For $\dot
E_{mag}/\dot E_{tot}\simgt 0.3$ most of the pressure of the inner
bubble is exerted close to the axis (see Figs. \ref{fig:beg} \&
\ref{fig:H}), which leads to the asymmetric evolution of the bounding
shell.}
\label{fig:grb}
\end{figure*}

In Figure \ref{fig:grb} we show the results of a series of simulations
for different fixed values of $\dot E_{mag}/\dot E_{tot}$, the ratio
of the Poynting flux to the total power injected by the magnetar. In
all of the calculations in Figure \ref{fig:grb}, we assume that the
total power supplied by the central source is given by $\dot E_{tot} =
10^{51} \, (t/1 \, {\rm s})^{-1} \, {\rm erg \, s^{-1}}$, which is a
reasonable approximation to the lower power solutions in Figure
\ref{fig:wind} (note that we neglect the possibility of early
injection of energy and start our simulation 1 second after core
bounce).  Note that in this case equal energy is supplied per decade
in time.

For values of $\dot E_{mag}/\dot E_{tot}\sim 0.1$, the pressure
distribution inside the bubble is relatively spherical
(Figs. \ref{fig:beg} \& \ref{fig:H}) and so the surrounding shell
becomes only modestly asymmetric.  Most of the energy supplied by the
central magnetar in this case is transferred to the surrounding SN shock
and host star.  Low magnetization bubbles of this kind would thus
likely produce a mildly asymmetric hyper-energetic SNe, but it appears
unlikely that the relativistic material supplied by the magnetar can
easily escape its host star.  For larger values of $\dot E_{mag}/\dot
E_{tot}$, the shell evolves more asymmetrically because most of the
pressure is exerted along the axis for magnetized bubbles.
By $\dot E_{mag}/\dot E_{tot} \simgt 0.3$, there is clear evidence for
a very elongated channel driven through the host star by the
anisotropic pressure of the central bubble. The shell reaches the outer
edge of the progenitor ($\sim 2\cdot 10^{10}$ cm) after $\approx
5-10$ sec.  At this point the ambient density drops to typical
circumstellar values, and the shell will rapidly blow out of the star.
The highly relativistic material contained in the interior bubble can
now flow relatively unimpeded out of the host star, forming a
relativistic jet; it is natural to associate these models with the
production of a GRB.

\begin{figure}
\resizebox{\hsize}{!}{\includegraphics[bb= 80 368 240 720, clip]{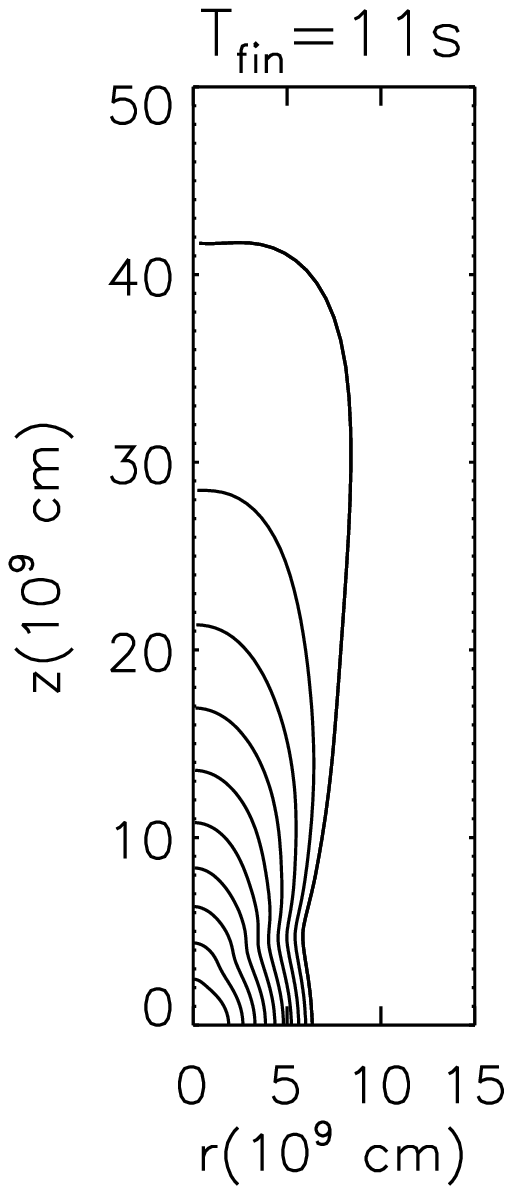}\includegraphics[bb= 80 368 240 720, clip]{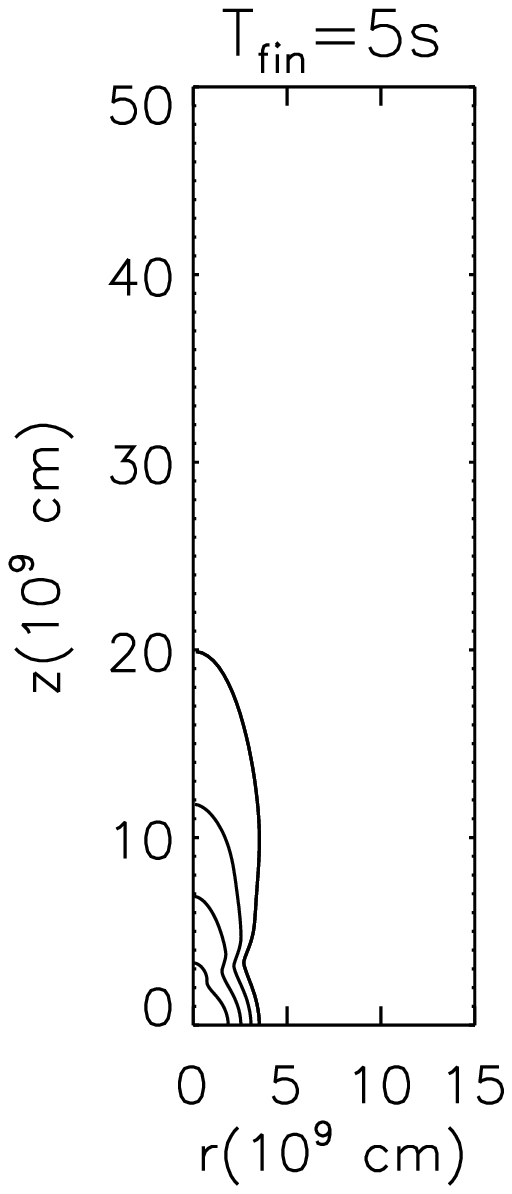}}
\caption{Evolution of a magnetized bubble inside a $35 M_\odot$
progenitor for a central source with a total spindown power of
$10^{52} (t/ 1 \, {\rm s})^{-2}$ erg s$^{-1}$; the initial time is 1s.
Contours represent the shape of the shell at 1s intervals.  {\it
Left:} $\dot E_{mag}/\dot E_{tot} = 0.2$ {\it Right:} $\dot
E_{mag}/\dot E_{tot} = 0.3$.}
\label{fig:evol}
\end{figure}

Figure~\ref{fig:evol} shows the evolution of the thin shell for a more
energetic, but more rapidly decaying, central source with $\dot
E_{tot} = 10^{52} \, (t/1 \, {\rm s})^{-2} \, {\rm erg \, s^{-1}}$,
which is an approximation to the higher power solutions in Figure
\ref{fig:wind}.  We consider $\dot E_{mag}/\dot E_{tot} = 0.2$ (left)
and $\dot E_{mag}/\dot E_{tot} = 0.3$ (right).  Note that in this
case, most of the energy is supplied to the bubble at early times and
so the evolution of the system is similar to the case of a magnetic
bomb with a fixed energy of $\sim 10^{52}$ ergs in the bubble.  The
evolution of the shell in Figure \ref{fig:evol} is qualitatively
similar to that of the lower power solutions shown in Figure
\ref{fig:grb}, although the bubble evolves more rapidly because of the
more energetic central source.  One consequence of this more rapid
evolution is that the shell velocity is closer to $c$, implying that
the assumption of magneto-hydrostatic equilibrium used to derive the
interior structure is likely to be less accurate than in the case of
the weaker power sources in Figure \ref{fig:grb}.

For PNSs with rotation periods longer than the values of $\approx 1-2$
ms considered in Figure \ref{fig:wind}, the energy injection rate will
be lower and approximately constant at early times because the
spindown time is longer than the Kelvin-Helmholz time of $\approx
10-100$ s.  To investigate this limit, we considered the evolution of
a bubble with a constant energy injection rate of $\dot E_{tot}
\approx 10^{50}$ erg/s. Elongation analogous to that shown in Figures
\ref{fig:grb} \& \ref{fig:evol} can be achieved, although somewhat
higher magnetization is required.  An asymmetry similar to the $\dot
E_{mag}/\dot E_{tot} = 0.2$ solution in Figure~\ref{fig:grb} requires
$\dot E_{mag}/\dot E_{tot} = 0.3$ for this lower $\dot E_{tot}$ and
takes a somewhat longer time $\sim 20$ sec to develop.  This example
highlights that lower power sources -- which can originate from more
modestly rotating PNSs -- can still lead to asymmetric bubbles because
the energy per unit solid angle along the pole is significant even for
modest $\dot E_{tot} \sim 10^{49}-10^{50} \, {\rm ergs \,
s^{-1}}$. Such sources may form asymmetric SN and, in some cases, very
long-duration GRBs or X-ray flashes.

An approximate analytic understanding of the late-time structure of
the shell shown in Figures \ref{fig:grb} \& \ref{fig:evol} can be
obtained by assuming that most of the energy is released inside the
bubble before it has time to expand significantly -- so that its shape
is still approximately spherical and adiabatic losses can be
neglected. In this case most of the pressure will reside in a region
along the axis whose opening angle is $\approx H/R$ and the
shell will expand rapidly along the axis within this solid angle.
Figure \ref{fig:H} gives the relationship between $H/R$ and
the magnetization of the bubble, which can be used to estimate the
opening angle of the resulting ``jet'' at late-times.  For example,
Figure \ref{fig:H} shows that $H/R \approx 0.1$ for $E_{mag}/E_{tot}
\approx 0.3-0.4$, which is reasonably consistent with the angular
collimation of $\sim 0.1$ rad in Figure \ref{fig:grb}.  It is also
worth noting that the high axial pressure produced by a magnetized
bubble with energy $E$ leads to an expansion along the axis that is
quantitatively similar to that produced by a jet with kinetic power
\begin{equation}
L_{jet}\simeq 7.5\cdot 10^{51} \, \theta \left(\frac{E}{10^{51} {\rm
erg}}\right)\left(\frac{10^9 {\rm cm}}{R}\right) {\rm erg \, s^{-1}},
\end{equation}
where we have assumed that the angular collimation $\theta \sim H/R$
and that the dimensionless axial pressure $\bar P \approx R/H$ (which
is reasonable for $H \simgt 0.03 \, R$; Figure \ref{fig:H}).

The results in Figures \ref{fig:grb} \& \ref{fig:evol} assume that
$\dot E_{mag}/\dot E_{tot}$ is independent of time.  This may be a
poor approximation given the strong evolution in the magnetization of
the wind as a function of time at small radii (Fig. \ref{fig:wind}).
Indeed, one might naively expect that $\dot E_{mag}/\dot E_{tot}$
would increase in time on a timescale of a few sec, given the
evolution of $\sigma_{LC}(t)$ for magnetar winds. If this is correct,
the magnetar-created bubble may initially impart its energy relatively
spherically, enhancing the energy of the outgoing SN shock (as in the
left panel of Figure \ref{fig:grb}). At late times, however, the
bubble will drive a jet out along the axis (as in the right-most panel
in Figure \ref{fig:grb}). To explore this scenario, we carried out a
series of simulations starting with a small initial value of $\sigma
\approx 0.05-0.1$ and then increasing $\sigma$ in time as
$\sigma_{LC}$ increases, up to the equipartition value of $\sigma
\approx 0.5$. As expected, in the first few seconds the evolution of
the shell was quite spherical, closely resembling the $\sigma=0.1$
panel of Fig.~\ref{fig:grb}, while at late times the shell evolves
into a more elongated structure analogous to the $\sigma=0.3-0.4$
panels of Fig.~\ref{fig:grb}. In this scenario, the different panels
in Fig.~\ref{fig:grb} qualitatively describe different snapshots in
time for the evolution of a shell driven by a wind with increasing
magnetization.  This suggests that the increasing magnetization of the
magnetar wind provides a way of tapping the spindown energy to
generate a mildly asymmetric hyper-energetic SN, while at the same
time creating an axial cavity along which relativistic material can
escape, powering a GRB.

Throughout this paper, we have considered only models in which the
progenitor has a spherical density profile, in order to understand the
importance of magnetic stresses on the asymmetric evolution of the
stellar wind bubble. However for rotating stars the density in the
polar region is expected to be smaller than at the equator. This can
also facilitate collimation.  A full investigation of the combined
collimation induced by magnetic stresses and a non-spherically
symmetric stellar profile is beyond the scope of this paper.  We have,
however, carried out a few preliminary calculations investigating the
two extremes: a low density polar channel whose opening angle is
either smaller or bigger than $H/R_{polar}$. Our results show that at
low magnetization the presence of a low density channel along the
rotation axis can lead to significant collimation (well in excess of
the mild asphericity shown in Figure \ref{fig:grb} at low
magnetization), while for higher magnetization the collimation is
primarily due to the magnetic stresses we have focused on in this
paper.
  
\section{Discussion}

In this paper we have calculated the evolution of a magnetized bubble
formed inside an exploding massive star.  Our motivation is to
understand the impact of a newly born millisecond magnetar on its
surrounding stellar envelope, and in particular to determine the
conditions under which magnetar birth can produce the collimated
outflows inferred from observations of long-duration GRBs.  

Neutron stars born with $P \sim 1$ ms and $B \sim 10^{15}-10^{16}$ G
lose $\sim 10^{51}-10^{52}$ ergs in $\sim 10-100$ sec in a magnetized
wind that becomes increasingly relativistic at late times (Figure
\ref{fig:wind}).  This energy forms a bubble of plasma and magnetic
field confined by the inertia of the surrounding star.  If the
material supplied to the bubble has low magnetization, $\dot E_{mag}/
\dot E_{tot} \simlt 0.1$, the bubble expands relatively spherically
(Figure \ref{fig:grb}) and most of the energy of the spinning down
neutron star is transferred to the surrounding SN shock, plausibly
powering a hyper-energetic SN, but probably not a GRB.  By contrast,
for more appreciable magnetization, $\dot E_{mag}/ \dot E_{tot} \simgt
0.3$, the bubble created by the magnetar rapidly becomes highly
elongated along the rotation axis of the system as a result of the 
internal pressure distribution (\S 2.1), forming a cavity out
of which the late-time relativistic wind from the magnetar can escape
(Figure \ref{fig:grb} \& \ref{fig:evol}).  We suggest that this is
plausible physical mechanism for forming collimated relativistic
outflows from newly formed millisecond magnetars.\footnote{Although we
have framed much of this paper in terms of the magnetar model for
GRBs, magnetized outflows from an accretion disk around a neutron star
or black hole would produce asymmetric magnetized bubbles akin to
those considered in this paper. Whether they would play a similar role
in driving collimated flows inside a supernova depends on the details
of the disks' mass and magnetic energy losses as a function of time,
as well as uncertainties in the disk winds' $\sigma$ similar to those
encountered in the neutron star winds considered here.}  

This mechanism works even if the host star is spherically symmetric.
In addition, even if most of the wind energy flux is concentrated in
the equatorial plane (as is expected for relativistic winds from
neutron stars), the surrounding bubble will still reach
magnetohydrostatic equilibrium and will thus elongate along the axis
due to magnetic stresses as we have described.  Finally, we note that
it is not necessary to wait until late times, when the magnetar wind
is relativistic, in order for the bubble to evolve highly
asymmetrically.  Collimation can be achieved in the early mass loaded
phase, provided only that the bubble contains a sufficient toroidal
magnetic field.  This mechanism may then operate in addition to
hoop-stress collimation of the non-relativistic flow (e.g.,
\citealt{me06}; \citealt{um06}).  This early time non-relativistic
phase cannot by itself produce a GRB, but can create a channel out of
which the later relativistic wind emerges.  Such a channel might also
provide boundary conditions conducive to the acceleration of the wind
and the conversion of magnetic energy into kinetic energy
\citep{kom07}.  Our calculations show that for the expected magnetar
energy loss rates, a collimated cavity is formed after $\sim 10$ sec
(Fig. \ref{fig:grb}).  At this point, magnetar winds have $\sigma_{LC}
\sim 100$ (Fig. \ref{fig:wind}), in the range required to account for
GRBs.

Because the birth rate of magnetars ($\sim 10 \%$ of neutron stars;
e.g., \citealt{kou94}) is significantly larger than the rate of GRBs
($\sim 0.1-1 \%$ of massive stellar deaths; e.g., \citealt{pod04}),
most magnetar births cannot produce standard long-duration GRBs.  This
is presumably either because an extended stellar envelope inhibits the
escape of a collimated outflow or because most magnetars are born
rotating more slowly than the millisecond rotators we have focused on
in this paper. For more modestly rotating PNSs, the asymmetric
expansion of a magnetized bubble could contribute to the inferred
asymmetry of many core-collapse supernovae (e.g., \citealt{wan01}).
In addition, a PNS with, e.g., $P \approx 4$ ms and $B \approx 3
\times 10^{15}$ G has a rotational energy of $\approx 10^{51}$ ergs
and a spindown time of $\approx 1$ day.  The birth of such a neutron
star would not produce a hyper-energetic SN or a canonical GRB.
However, if the bubble created by the magnetar is sufficiently
magnetized, it would evolve asymmetrically in a manner similar to the
calculations shown in Figures \ref{fig:grb} \& \ref{fig:evol}.  This
could produce a long-duration transient analogous to the X-ray flash
060218 associated with SN 2006aj (\citealt{maz06b}; \citealt{sod06};
we should note, however, that many X-ray flashes may have lower
inferred energies because of viewing angle effects rather than being
intrinsically less energetic events; e.g., \citep{granot}.  The
remnant Cass A, with its strong jet/counter-jet morphology (e.g.,
\citealt{hwa04}), may be an example of an asymmetric explosion driven
and shaped by a magnetized wind accompanying magnetar birth.  Indeed,
\citet{cha01} suggested that the central X-ray point source in Cass A
is a magnetar.

The thin-shell calculations described in this paper assume that the
magnetar wind expands into an initially spherical cavity created by an
outgoing SN shock.  This requires that the spindown time of the
magnetar is at least somewhat longer than the time required to
initiate the stellar explosion (i.e., $\simgt 1-2$ sec). Our
assumption of a ``successful'' SN explosion does not, of course,
preclude that the explosion itself is magneto-centrifugally driven, as
in the force-free model for the collimated explosion of a star by a
newly-formed magnetar in an otherwise ``failed'' SN (e.g.,
\citealt{ost71} or \citealt{um07}).  However, one interesting problem
not addressed by our calculations is the spindown of the magnetar and
the evolution of its surrounding bubble if the initial explosion is
primarily bipolar (see, e.g., the simulations of \citealt{moi06} and
\citealt{burrows07}).  Late-time collimation of relativistic material
in this context may be modified by the large inertia of the accreting
stellar envelope (or fallback material) in the equator of the star
(see also the related arguments of \citealt{um07}). In addition, it is
worth noting that if the outflow always has high magnetization, our
calculations suggest that, because most of the pressure will be
exerted along the axis, there could be a collimated GRB but no
associated equatorial explosion.  This could account for the recently
discovered supernova-less GRBs \citep{fyn06}.

One of the uncertainties associated with our calculations is that the
magnetization of the material supplied to the surrounding bubble is
difficult to calculate. Magnetic energy has to be supplied to the
bubble relatively rapidly, with $ \sigma = \dot{E}_{mag}
/\dot{E}_{tot} \simgt 0.2$ at the termination shock in our models that
show significant collimation.  Observations of PWNe suggest quite low
$ \sigma \sim 0.01$ at the termination shock, which would imply that
there is insufficient time to build up the anisotropic magnetic stress
needed to drive aspherical expansion of the surrounding stellar
envelope. However, we suspect that the confined bubbles around newly
formed magnetars will have higher magnetization at their termination
shocks than has been inferred in PWNe and in pulsar-Be star
binaries. This is because the distance to the termination shock is
only $\sim 10$ light cylinder radii in our problem, relative to $>
10^4$ light cylinder radii in the systems where we have direct
observational constraints.  As a result, there is less time for the
magnetic field in the wind to dissipate, plausibly leading to higher
magnetization.

All of the calculations described in this paper are based on the
thin-shell approximation.  This model is useful for demonstrating that
magnetar birth can produce conditions conducive to the formation of a
collimated outflow that can emerge out of the host star.  However,
modeling this process in detail is beyond the scope of the present
simplified calculations, and will require full relativistic MHD
simulations.  Indeed, it is our intention to use the results of the
present paper as a guide for more realistic simulations. Such
calculations are necessary to determine the fraction of the spindown
energy that goes into a relatively spherical explosion of the host
star relative to the energy that flows out of the collimated cavity.
Quantifying this is important for understanding the conditions under
which magnetar birth might produce {\it both} a hyper-energetic SN and
a GRB, as is observed (e.g., \citealt{woo06}).  We have speculated in
\S4 that this can occur if the surrounding bubble becomes
progressively more magnetized as the magnetar spins down, but
multi-dimensional simulations are needed to assess this (and to
understand if it is a necessary condition).  Multi-dimensional
simulations will also allow a more detailed study of how the late-time
relativistic outflow emerges from within the earlier non-relativistic
wind and the different observational signatures associated with each
phase of the outflow (analogous to studies of jets emerging from a
host star; e.g., \citealt{MLB07}).  They also allow investigation of
the stability of the confining envelope, which is subject to possible
Rayleigh-Taylor fragmentation, since it is accelerated by the light
weight bubble (e.g., \citealt{arons03}), an effect likely to be of
substantial significance at higher energy injection rates. Such
instabilities could be important for understanding the short timescale
variability observed in GRBs, as could intrinsic time variability in
the magnetar wind, driven by, e.g., reconnection in the equatorial
current sheet \citep{me06}.  Variability might also originate inside
the bubble due to the dynamics of the interaction of the wind with the
progenitor (as is seen in PWNe simulations; e.g.,
\citealt{ldz04}). Finally, in addition to the asymmetric expansion of
the entire bubble discussed here, observations and simulations of PWNe
also reveal a moderately relativistic axial ``jet'' within the nebula
itself (e.g., \citealt{weiss00,ldz04}); this may be dynamically
important for proto-magnetar bubbles as well.

\section*{Acknowledgments}
N.B. was supported by NASA through Hubble Fellowship grant
HST-HF-01193.01-A, awarded by the Space Telescope Science Institute,
which is operated by the Association of Universities for Research in
Astronomy, Inc., for NASA, under contract NAS 5-26555.  EQ and BDM
were supported by the David and Lucile Packard Foundation and a NASA
GSRP Fellowship to BDM. JA's research has
been supported by  NSF grant AST-0507813, NASA grant
NNG06GI08G, and DOE grant DE-FC02-06ER41453, all at UC Berkeley; by
the Department of Energy contract to the Stanford Linear Accelerator
Center no. DE-AC3-76SF00515; and by the taxpayers of California.
We thank A.~Heger for making massive stellar progenitor models 
available online.

\begin{appendix}

\section{Self-Similar Solutions}

It is well known that the 1D thin shell equations admit self-similar
solutions, where the time dependence can be factored out.
\citet{chev94} have shown that equations (\ref{eq:1})-(\ref{eq:6}) can
be solved in the simplified case of a wind blowing inside a wind, when
the shell expands at a constant speed. In this case it is possible to
reduce equations (\ref{eq:1})-(\ref{eq:6}) to a system of ordinary
differential equations.  Here we extend their treatment by considering
the expansion of a hot magnetized bubble inside either a stationary
medium (as in the case of the ISM or the roughly static envelope of a
progenitor star) or freely expanding ejecta (as in the case of
PWNe). We derive the conditions under which self-similarity applies
and discuss how the elongation of the bubble scales with its magnetic
energy content. We limit our investigation to the case of an isotropic
outer medium, even though self-similarity holds in the more general
case of fixed latitudinal dependence (\citealt{kah85},
\citealt{cam03}).

If one assumes a cold unmagnetized outer medium with a power law
density profile $\rho_o\propto r^\alpha$, that $p_n\propto t^\chi$ and
that $R_{polar}/H$ is constant, where $R_{polar}$ is the radius of the
shell along the axis, one can search for a solution of the form
\begin{eqnarray}
R=R(\theta)t^\beta\nonumber\\
v_\bot=v_\bot(\theta)t^\delta\nonumber\\
v_\|=v_\|(\theta)t^\delta\nonumber\\
\sigma=\sigma(\theta)t^\eta\nonumber
\end{eqnarray}
Taking $\delta=\beta-1$, $\eta=\beta+\alpha\beta$, and
$\chi=2\beta+\alpha\beta-2$, the time dependence can be eliminated and
the problem reduces to a system of ordinary differential equation in
the $\theta$ direction. Neglecting the outer pressure and magnetic
field, and the inner ram pressure, and assuming a stationary outer
medium (a similar derivation can be obtained for freely expanding
ejecta), equations (\ref{eq:1})-(\ref{eq:6}) can be written as
\begin{eqnarray}
\tan{\xi}=-\frac{1}{R}\frac{\partial R}{\partial \theta},\\
\beta \sigma=\rho_o(\beta R\cos{\xi})-2\beta\sigma-\nonumber\\
\frac{\cos{\xi}\partial}{\sin{\theta}R^2\partial \theta}\left[R\sin{\theta}\sigma(v_{\|}-\beta R\sin{\xi})\right]\\
 \beta (\beta -1) \sigma R\cos{\xi}=
-\rho_o\beta^2R^2\cos^2{\xi}+(p_i+B_i^2/8\pi)-\nonumber\\
\sigma [v_\|+\beta R\sin{\xi}]\frac{\cos{\xi}}{R}\left[\frac{\partial }{\partial \theta}(\beta R \cos{\xi})-v_\|\left(1+\frac{\partial\xi}{\partial\theta}\right)\right]\\
(\beta -1) \sigma v_\|=
-\rho_o\beta R\cos{\xi}v_\|-\nonumber\\
\sigma [v_\|+\beta R\sin{\xi}]\frac{\cos{\xi}}{R}\left[\frac{\partial }{\partial \theta}(v_\|)+\beta R\cos{\xi}\left(1+\frac{\partial\xi}{\partial\theta}\right)\right],
\end{eqnarray}
where the variables are now only function of $\theta$, and
$p_i+B_i^2/8\pi=p_nf(R\sin{\theta}/H)$. Note that these equations are
not easy to write in non-dimensional form because, as we will discuss
below, the characteristic length-scale is an eigenvalue of the
problem. Only in the case $\alpha=-2$ is the solution relatively
straightforward (\citealt{chev94}).  

To check the validity of the self-similar solution one must verify a
posteriori that $R_{polar}/H$ does not change in time and that the
pressure $p_n$ scales as a power law.  We first consider the 
scale height $H$. As discussed in \S \ref{bottle}, the evolution of the
bubble can be described as a series of infinitesimal adiabatic
expansions followed by energy injection at fixed volume. It is easy to
show that, in the relativistic case, when the adiabatic index is
$4/3$, entropy and magnetic flux conservation imply that the ratio
$R_{polar}/H$ is indeed constant during self-similar expansion.  On
the other hand, as long as the ratio of the Poynting flux to the total
luminosity, $\dot E_{mag}/\dot E_{tot}$, is equal to the ratio of the
magnetic energy to total energy in the bubble $E_{mag}/E_{tot}$, the
height scale $H$ is constant, at fixed shape.
 
Despite the fact that the thin-shell equations conserve only momentum,
it possible to conserve energy as well by requiring the correct
temporal scaling, in which case the total energy in the system
increases only because of injection ($\dot E_{tot}$). The total energy
in the shell is:
\begin{equation}
\int_{S}\frac{1}{2}\sigma(v_\|^2 +v_\bot^2)dS=E_{sh}t^{5\beta+\alpha\beta -2},
\end{equation}
where $S$ is the shell surface and $E_{sh}$ is a constant that depends
on the shell structure. If the ratio $R_{polar}/H$ is constant then
the total energy in the bubble is
\begin{equation}
\int_{V}e(r,z)dv=p_nKt^{3\beta},
\end{equation}
where $V$ is the volume of the bubble, $K$ is a constant depending on
the shell shape and the ratio $R_{polar}/H$, and $e$ is the energy
density in the bubble. If $p_n\propto t^\chi$ then the total internal
energy scales as the shell energy. Thus one can write the total energy
in the system as
\begin{equation}
E_{tot}(t)=\int_o^t\dot E_{tot}dt\propto t^{5\beta+\alpha\beta -2}.
\label{eq:inj}
\end{equation}
This relation shows that as long as $\dot E_{tot}$ is a power-law in
time, self similar solutions are self-consistent. Equation
(\ref{eq:inj}) together with the ratio $\dot E_{mag}/\dot E_{tot}$
determines how the self-similar shape is related to the injection
properties, namely the total luminosity and Poynting flux. Once
$\alpha$ and $R_{polar}/H$ are given one can search for a solution of
equations (A1)-(A4).

\subsubsection{Numerical Methods}

In the case studied by \citet{chev94}, the shell moves at a constant
speed, $\beta=1$, and the system of ordinary differential equations
can be easily solved as an initial value problem, assuming regular
conditions on the axis. Given that the equations are singular on the
axis, it is necessary to expand them in Taylor series and solve for
the coefficients. In more general cases, the shell is subject to
acceleration and $\beta\neq 1$. This changes the nature of the
equations and makes the problem much harder to solve. If one tries to
expand the equations in a Taylor series on the axis, which works for
the case $\beta=1$, one finds that terms of order ${\cal O}(\theta^n)$
 depend on terms of order ${\cal O}(\theta^{n+1})$, 
and the series is open. This is typical of two-point boundary
value problems (eigenvalue problems) where conditions at one boundary
are not sufficient. However it is not obvious where and what kind of
additional boundary conditions one should impose. One might enforce
regularity on the equator, either by assuming a smooth shape or zero
parallel velocity. However the solutions of \citet{chev94} show that
discontinuities at the equator might arise, especially for larger
value of $R_{polar}/H$. We have chosen to impose that the tangential
transport velocity should be $0$ on the equator to avoid the formation
of an equatorial ring where matter can accumulate indefinitely, but
this does not preclude other possible boundary conditions.

We tried several methods for solving the steady state angular
equations given by (A1)-(A4).  These included standard shooting
techniques, using both explicit ODE solvers and an implicit
integrator, standard relaxation methods, and a Fourier expansion of
the angular equations.  For a variety of reasons, we find that none of
these methods was satisfactory for finding solutions, aside from the
relatively spherical limit when $R_{polar}/H \simlt 1$.  Instead, we
found that the most successful method was to use the time dependent
equations (eqs. [\ref{eq:1}]-[\ref{eq:6}]) to evolve the system
forward in time, imposing a fixed value for $R_{polar}/H$ and a
pressure $p_n\propto t^\chi$, where the exponent $\chi$ is the one
expected for the self-similar solution.  Self-similarity is then
reached at late times after the initial transient dies away; we have
verified that the shell at late times has converged to a self-similar
shape, with only a few percent error in radius.  It is, however, known
that accelerated shells are subject to corrugational instabilities
(the thin-shell instability).  We have included an artificial
viscosity to suppress this instability and focus on the overall
evolution of the bubble.  Instabilities may indeed be important for
the evolution of the shell but we leave a study of them to future
work. In each case, we have determined a posteriori that the chosen
value of viscosity was small enough not to affect the shape of the
bubble. This method for determining the self-similar solution worked
up to $R_{polar}/H \sim 10$. More elongated bubbles were difficult to
investigate unless one started with initial conditions very close to
the desired self-similar solution.  Otherwise the deviation from the
correct self-similar solution was effectively a large-amplitude
large-scale initial perturbation; in this case, we found that the
corrugational instability could not be suppressed without increasing
the viscosity to a point where it modified the overall shape of the
bubble.

\begin{figure*}
\hspace*{-2cm}\includegraphics[width=12cm]{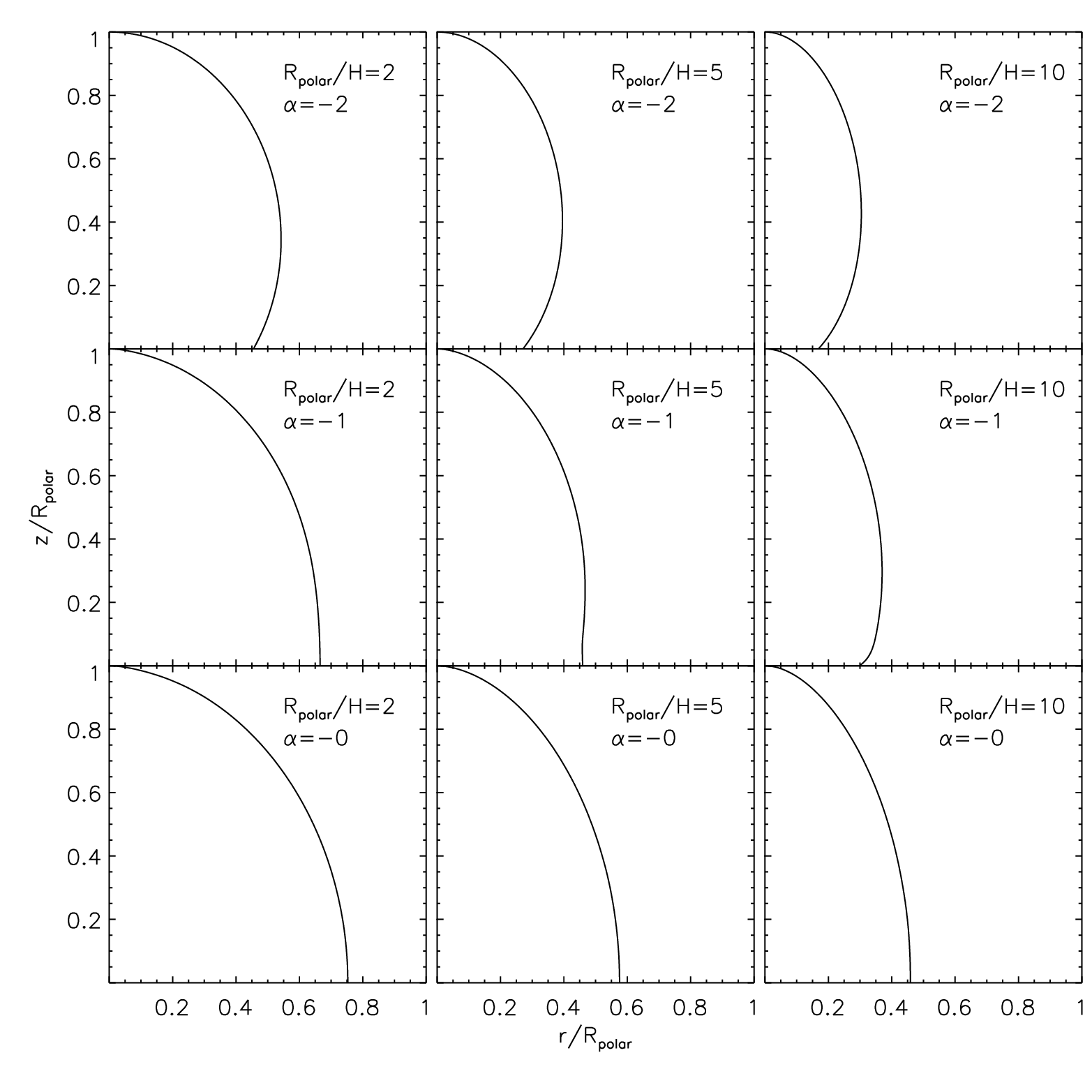}
\caption{Self-similar shapes for magnetically driven bubbles expanding
into a static ambient medium with $\rho \propto r^{\alpha}$. Results
are shown for three different values of $R_{polar}/H$.  Results for
$\alpha=-2$ are based on the model of \citet{chev94} while for $\alpha
= 0$ and $-1$, they are based on time dependent thin-shell simulations
that reach self-similarity at late times.}
\label{fig:shape}
\end{figure*}

\subsubsection{Results}

In Figure \ref{fig:shape} we plot the shape of the shell for three
different values of $R_{polar}/H$, and for $\alpha=0,-1,-2$ in the
case of a constant injection luminosity ($\beta = 3/(5+\alpha)$).  
 For $\alpha = 0$ and
$-1$, the solutions are found by the methods described in the previous
subsection and are restricted to $R_{polar}/H \simlt 10$.  The $\alpha
= -2$ case is that of \citet{chev94} for which the angular equations
(A1)-(A4) can be straightforwardly integrated.  In this case, we
compute solutions up to $R_{polar}/H \approx 100$, for comparison to
the time-dependent solutions in the main text.

Figure \ref{fig:shape} shows that the elongation of the bubble
increases with increasing $R_{polar}/H$, and for smaller values of
$\alpha$. In the case of a constant density outer medium, the shell
appears to be regular both at the pole and at the equator, while in
the case $\alpha=-1$ and $-2$ for large values of $R_{polar}/H$, a
cusp is formed in the equatorial plane, where matter tends to
accumulate. We have verified that all quantities scale according to
self similarity.

Unlike in the case studied by \citet{chev94}, in which the surface
density monotonically increases from the pole to the equator, for
$\alpha=0$ and small values of $R_{polar}/H$ we find that the surface
density reaches a maximum at intermediate latitudes.

\begin{figure}
\resizebox{\hsize}{!}{\includegraphics[bb=70 375 570 720, clip]{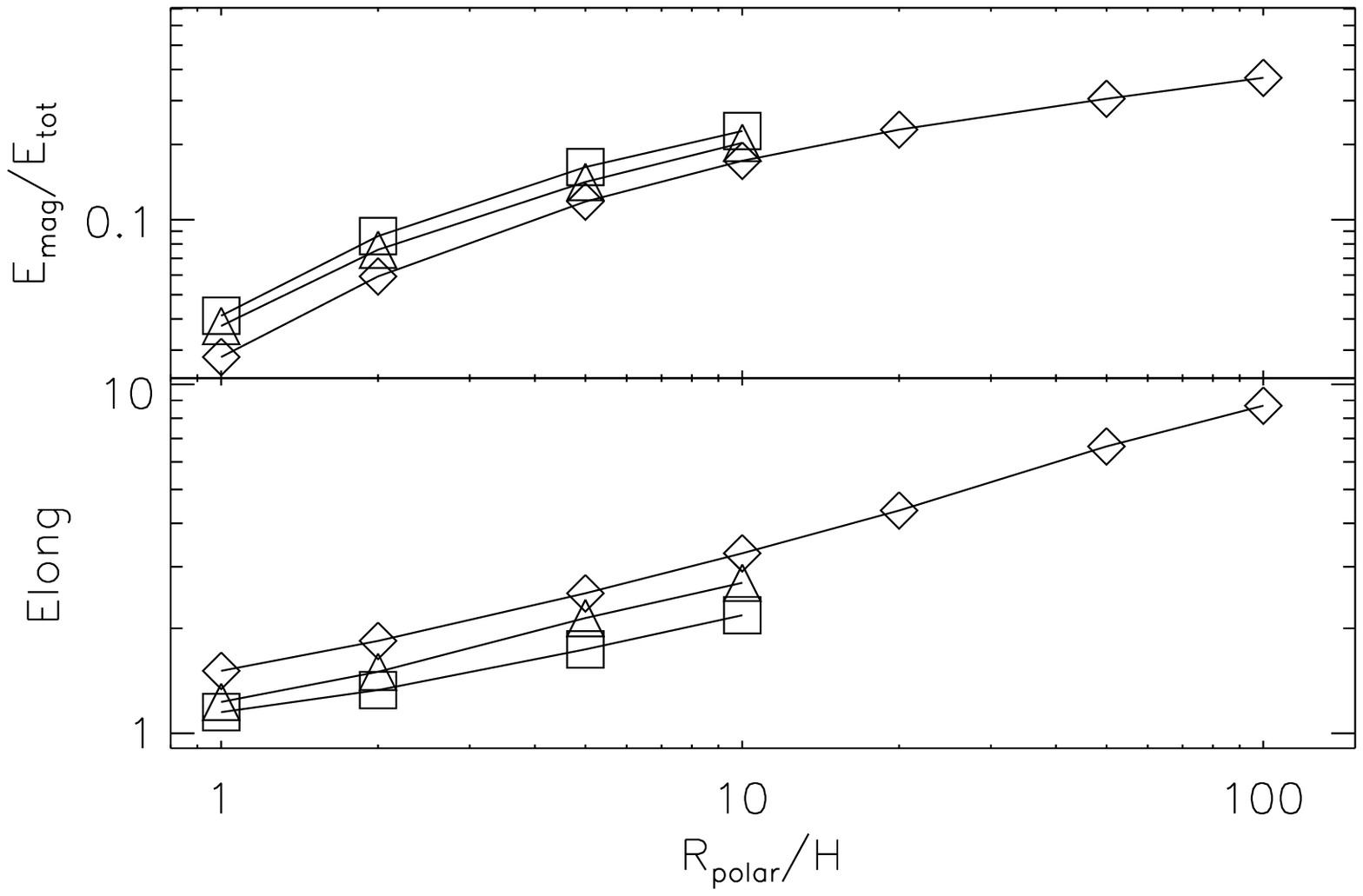}}
\caption{Properties of magnetized bubbles expanding into a stationary
medium with a power law density profile with $\rho \propto
r^{\alpha}$, for $\alpha=0$ (squares), $\alpha=-1$ (triangles), and
$\alpha=-2$ (diamonds).  The $\alpha = -2$ results are based on
solving the steady state angular equations (A1)-(A4), which can be
solved for larger values of $R_{polar}/H$ than the $\alpha = 0$ and
$-1$ cases, which are based on solving the time dependent thin shell
equations.  $E_{mag}/E_{tot}$ is the
ratio between the magnetic and total energy in the bubble. The
elongation is defined as the ratio between the length of the bubble
along the axis and the maximum extent of the bubble away from the
axis.}
\label{fig:andam1}
\end{figure}

Figure \ref{fig:andam1} shows a summary of the results of our
simulations for the case of constant injection power into a static
outer medium (typical for wind bubbles): shown are the ratio of the
magnetic energy to the total energy and the elongation of the bubble,
as functions of $R_{polar}/H$.  The elongation scales approximately as
$\sqrt{R_{polar}/H}$ and, as expected, is larger for steeper density
profiles. The dependence on the density profile of the outer medium
is, however, quite weak, with less than a factor of 2 difference
between $\alpha=0$ and $\alpha=-2$.  For the magnetar problem
considered in this paper, this suggests that the outer density
structure of the progenitor star should not significantly effect the
collimation of the bubble (this is consistent with the fact that the
asymmetry of the bubble was only a weak function of the progenitor
star, as discussed in \S \ref{bottle}).  Figure \ref{fig:andam1} also
shows that in all cases the ratio $E_{mag}/E_{tot}$ inside the bubble
is relatively small, and tends to be smaller for smaller values of
$\alpha$. This demonstrates that strong elongation  does not
require magnetically dominated bubbles, which is consistent with our
result for the evolution of magnetized bubbles inside GRB progenitors
in \S \ref{bottle}.

\begin{figure}
\resizebox{\hsize}{!}{\includegraphics[bb=70 375 570 720, clip]{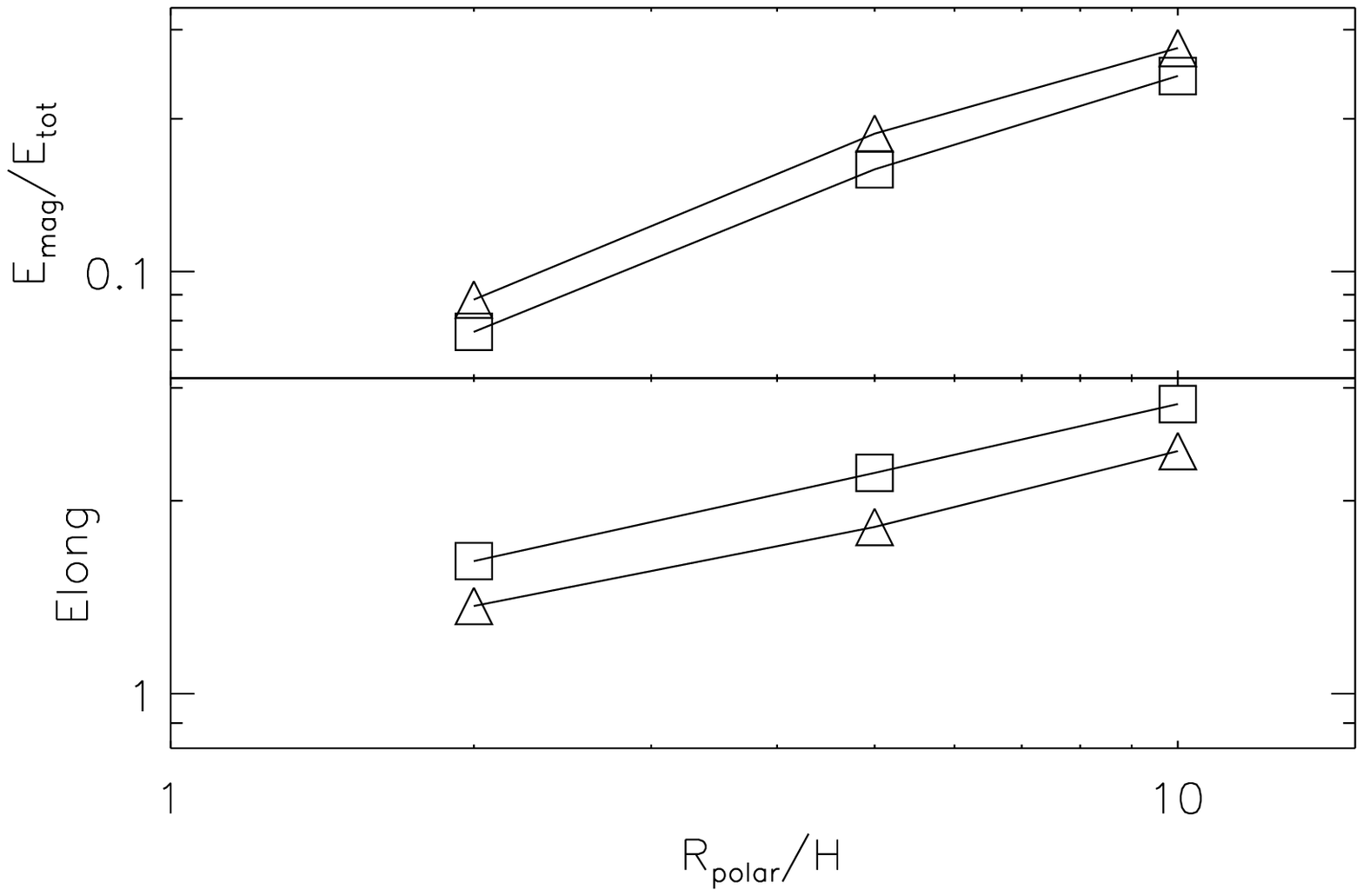}}
\caption{Same as Figure \ref{fig:andam1} but for the case of expansion
into the freely expanding ejecta of a supernova remnant: $v_o\propto
R/t$, $\rho\propto r^{\alpha}t^{-\alpha-3}$; results for $\alpha=0$
(squares) and $\alpha=-1$ (triangles) are shown.}
\label{fig:andam2}
\end{figure}

We also carried out simulations for a bubble inside freely-expanding
ejecta with velocity $v_o\propto R/t$ and density $\rho\propto
r^{\alpha}t^{-\alpha-3}$.  We find that for $\alpha=-2$, the
time-dependent solution never reaches a self-similar solution, but
instead the evolution always depends on the initial conditions; for
this reason, we do not show solutions for $\alpha = -2$.  This because
in the case $\alpha=-2$ if one assumes the expected power law temporal
evolution for the internal pressure, it is easy to show that the ratio
between the inner pressure and the ram pressure of the outer medium is
constant, and does not depend on the radial evolution. In this sense,
the bubble never relaxes to self-similarity.  The case of
freely-expanding ejecta is relevant for a central source which is not
energetic enough to produce a bubble which effects the dynamics of the
stellar explosion on short time-scales, but instead creates an
energetic PWN inside the expanding SN ejecta.  This could be relevant
to young PWN generally and also to lower energy transients such as
X-ray flashes (see \S 5).  The results for the elongation of the
bubble and its dependence on the magnetization are shown in Figure
\ref{fig:andam2}; they are reasonably similar to the results shown in
Figure \ref{fig:andam1}.

\end{appendix}

\label{lastpage}

\end{document}